Title page with author details

11,614 words in the main text

246 words in the abstract

123 references

13 tables and 6 figures in the main text

5 tables and 1 figure in the Appendix

# Wait or cross? Understanding the influence of behavioral tendency, trust, and risk perception on pedestrian gap-acceptance of automated truck platoons

Yun Ye[1,2,3], Yuan Che[1,2], Haoyang Liang[4], Yingheng Zhang[1,2], Pengpeng Xu[5†]

[1] Faculty of Maritime and Transportation, Ningbo University, Ningbo, China

[2] Collaborative Innovation Center of Modern Urban Traffic Technologies, Southeast University, Nanjing, China

[3] Centre for Transport Engineering and Modelling, Department of Civil and Environmental Engineering, Imperial College London, London, UK

[4] College of Transportation Engineering and the Key Laboratory of Road and Traffic Engineering, Ministry of Education, Tongji University, Shanghai, China

[5] School of Traffic and Transportation Engineering, Central South University, Changsha, China

† **Correspondence to**: pengpengxu@yeah.net

**Funding:** This research was supported by grants from the National Natural Science Foundation of China (Project No. 72501150, 52302433), Zhejiang Provincial Natural Science Foundation of China (Grant No. LQN25E080011), Ningbo Natural Science Foundation (Grant No. 2024J440), Natural Science Foundation of Guangdong Province, China (Grant No. 2024A1515011578), and National "111" Centre on Safety and Intelligent Operation of Sea Bridges (Project No. D21013). The funders had no role in the study design, data collection and processing, manuscript preparation, or decision to publish.

# Wait or cross? Understanding the influence of behavioral tendency, trust, and risk perception on pedestrian gap-acceptance of automated truck platoons

## ABSTRACT

Although automated trucks have the potential to improve freight efficiency, reduce costs, and address driver shortages, organizing two or more trucks in a convoy has raised considerable concerns for pedestrian safety. This study conducted a controlled experiment to examine the influence of behavioral tendency, trust, and risk perception on pedestrian intention to cross in front of an automated truck platoon. A total of 603 subjects participated in the virtual reality video-based questionnaire survey. By fusing the merits of structural equation modeling and artificial neural networks, a two-stage, hybrid model was developed to examine complex relationships between latent variables and gap-acceptance behaviors. Our results indicated that subjects watched an average of five vehicle gaps before starting crossing and the average time gap accepted was about 5.35 seconds. Risk perception not only played the most dominant role in shaping pedestrian crossing decisions, but also served as the strong bone, mediating the effects of behavioral tendency and trust on gap-acceptance. Participants who frequently violated traffic rules were more likely to accept a smaller time gap, while those who showed positive behaviors to other road users tended to wait for a larger time gap. Participants who often committed errors, showed aggressive behaviors, and held greater trust in the safety of automated trucks generally reported a lower level of risk for road-crossing in front of automated truck platoons. Built on these findings, a range of tailored countermeasures were proposed to ensure safer and smother interactions between pedestrians and automated truck platoons.

*Keywords*: Autonomous trucks; virtual reality; pedestrian gap acceptance; risk perception; behavioral tendency.

## 1. Introduction

As the backbone of freight transportation, trucks play an indispensable role in global supply chain. In China, trucking accounted for 73.7% of the national freight traffic in 2023, yielding an 8.7% increase from 2022 (MTPRC, 2024). Despite the critical role of trucking, there is a growing shortage of truck drivers. In 2022, the shortage rate of professional truck drivers reached as high as 31.60% in China, substantially higher than that of passenger vehicle drivers (CRTA, 2023). Fortunately, the emerging automated driving technology presents a potential avenue to alleviate truck driver shortages.

The penetration of automated vehicles (AVs), however, has raised considerable public concerns, particularly about the safety of vulnerable road users (Das, 2021; Tran et al., 2021; Rezwana and Lownes, 2024; Li et al., 2025a). Some pedestrians feel anxious and uncertain when interacting with AVs, as these vehicles cannot communicate intentions through implicit, non-verbal cues like eye contact, head nodding, and hand gestures, leading to hesitation and under-confidence when making crossing decisions (Nuñez Velasco et al., 2019). Others speculate that AVs will always yield and thereby take aggressive actions, which induce severe, unexpected conflicts (Rodríguez Palmeiro et al., 2018).

Unlike passenger cars, the characteristics of trucks (e.g., large size, heavy weight, and constrained maneuverability) pose unique risks to pedestrians (He et al., 2024). In-depth crash investigation indicates that the mortality rate for pedestrians stuck by trucks is about two times higher than that of passenger cars (Roudsari et al., 2004; Schubert et al., 2023). In this regard, organizing two or more trucks in a convoy via connected and automated driving technologies promises to revolutionize the freight industry by enhanced road safety, improved fuel efficiency, reduced carbon emissions, and alleviated traffic congestion (Tsugawa et al., 2016; Noruzoliaee and Zhou, 2021; Lourenço et al., 2024; Li et al., 2025b; Lin et al., 2025). When interacting with pedestrians, a conservative platooning strategy that prioritizes pedestrian safety may compromise efficiency, as all the following trucks must stop when the leading truck does. Frequent and sudden decelerations, particularly for the heavily loaded trucks, are likely to augment safety hazards and undermine the benefits of truck platoons. Meanwhile, maintaining short gaps within the platoon to prevent nearby traffic from cutting in, however, may prompt pedestrians to cross even without a safe gap. The situation becomes far more complicated at the early stage of truck platooning deployment, because intentions may not be explicitly conveyed in absence of vehicle-to-pedestrian communications. Although pedestrian interactions with human-driven vehicles (Holland and Hill, 2007; Lobjois and Cavallo, 2007; 2009; Lobjois et al., 2013; Petzoldt et al., 2014; Alver and Onelcin, 2018; Avinash et al., 2019; Vasudevan et al., 2020; Salducco et al., 2022; Osorio-García et al., 2023) and automated vehicles (Deb et al., 2018; Rodríguez Palmeiro et al., 2018; Dey et al., 2019; Nuñez Velasco et al., 2019; Woodman et al., 2019; Rad et al., 2020; Camara et al., 2021; Kalatian and Farooq, 2021; Zhao et al., 2022; Feng et al., 2024) have been extensively studied over the past two decades, there are surprisingly limited studies on gap-acceptance of automated truck platoons (Li et al., 2025b; Lin et al., 2025). An in-depth understanding of pedestrian behaviors interacting with truck platooning not only supports the formulation of evidence-based measures to enhance pedestrian safety, trust, and receptivity, but also informs the design of socially aware organization algorithms for truck platoons, ultimately fostering safer and more cooperative interactions between humans and automated systems in real world environment.

To fill this gap, this study conducts a controlled experiment to examine the influence of behavioral tendency, trust, and risk perception on pedestrian intention to cross in front of an automated truck platoon. The integration of virtual reality (VR) simulation with stated-preference survey allows us to collect a semantically rich, realistic dataset on both objective and subjective measures from a relatively large sample size in a safe, reproducible, controllable, and cost-effective manner, while the development of a hybrid structural equation modeling and artificial neural networks (SEM-ANN) model enables to untangle the complex relationship between latent variables and gap-acceptance behaviors. Specifically, the main contributions of this study are summarized as follows:

- To the best of knowledge, our study is among the first to investigate the pedestrian gap-acceptance of autonomous truck platoons via leveraging the pragmatic advantages of VR simulation and self-reported questionnaire survey.
- By harnessing the well-established Pedestrian Behavior Questionnaire (PBQ; Deb et al., 2017a) to quantify the frequency of risky behaviors conducted in daily life, we advance the characterization of pedestrian heterogeneity in behavioral tendency and crossing decision, a dimension underexplored in previous studies on pedestrian–AV interactions.
- To fuse the merits of SEM and ANN, a hybrid, two-stage model is established to uncover the interconnection between latent variables and pedestrian intention to cross. Such practice provides a promising, flexible tool for analysts when modeling intricate relationships among latent variables, which is frequently encountered in traffic psychology and travel behavior studies.

The remainder of the paper is structured as follows. Section 2 presents the literature review. Section 3 outlines the research methodology, including data collection, variable measurement, and the SEM-ANN approach. Section 4 presents the modeling results, which are discussed in Section 5. Section 6 concludes the paper and highlights the directions for future work.

## 2. Literature Review

### 2.1 Automated truck platoon

Since its introduction in the middle 1990s, linking multiple trucks in a platoon using vehicle automation and communication techniques has attracted considerable research interest. Earlier studies focus primarily on the organization, safe control, energy-saving, and technology acceptance of truck platooning. Numerical simulation experiments and field trials have shown that truck platooning with an inner gap of about 10 m could save 10%–15% of fuel consumption (Alam et al., 2010). In addition to the energy-saving, by optimizing the lateral position of truck platoon via a centralized control strategy, the pavement damage accumulation and life-cycle costs could also be reduced up to 50% (Gungor and Al-Qadi, 2020). Beyond the overwhelming benefits of platooning technology to freight industry, its adaptation ultimately depends on the receptivity from decision-makers, truck drivers, and general public who will necessarily interact with truck platoons on the road. After reviewing a total of 35 articles, Lourenço et al. (2024) summarized the representation of various stakeholders on truck platooning acceptance and found that decision-makers were more concerned about the operation efficiency, while truck and peripheral drivers highlighted the safety and reliability of platooning technology.

As on-road testing continues, researchers have begun to investigate the barrier effect of truck platoons on nearby traffic. Sultana and Hassan (2024) explored whether human drivers behaved differently when interacting with recognizable connected and automated truck platoons by conducting a controlled driving simulation experiment with 42 participants. Significant improvement in merging behaviors was reported in the presence of recognizable truck platoons, with the time-to-collision being 20% higher and the standard deviation of speed and acceleration being around 32% and 29% lower, respectively. More recently, Li et al. (2025b) conducted a driving simulation experiment with 40 participants to examine the driver lane-changing and off-ramp behaviors near highway exists under the influence of a platoon with three trucks. The results indicated that a larger platoon gap increased the merging rate and driver confidence in overtaking, while a smaller gap resulted in a higher proportion of late lane changes. Likewise, Lin et al. (2025) recruited 38 participants in a high-fidelity driving simulation experiment and investigated the influence of platoon speed (80 km/h and 100 km/h), size (three trucks and five trucks), inner gap (5 m and 25 m), and the presence of a leading vehicle on overtaking behaviors of drivers. Their study found that a higher platoon speed raised the overtaking difficulty and posed more drivers to take risky actions, while a smaller inner gap significantly increased the likelihood of successful overtaking maneuvers.

Although the aforementioned studies have shed insightful light on how human drivers adapt their lane-changing, overtaking, merging, and diverging behaviors when interacting with the automated truck platoon on freeways, no studies, to date, have particularly analyzed pedestrian behaviors in interactions with truck platoons. Gaining a thorough understanding of the factors influencing pedestrian gap-acceptance of automated truck platoons not only enhances the safety and efficiency, but also promotes the upgrade and acceptance of this foreseeable technology.

**2.2 Pedestrian gap-acceptance behavior**

When crossing the road, the decision made by a pedestrian to accept or reject a specific gap between two consecutive approaching vehicles is regarded as the gap-acceptance behavior. One variable of interest is the critical gap, which is defined as the time in seconds that a pedestrian would not attempt to cross (Theofilatos et al., 2021). Objectively, how safe a specific gap is mainly dependent on the time needed for a pedestrian to step into the lane and reach the far side of the street (i.e., crossing duration) and how long it would take the approaching vehicle to arrive (i.e., time-to-arrival; Ahmed et al., 2024). Safety margin (also known as the post-encroachment time) can then be defined as the difference between the time a pedestrian completes crossing and the time the next oncoming vehicle arrives at the conflict line (Alver and Onelcin, 2018). Quantifying the size of accepted gaps and safety margins, together with uncovering the underlying mechanism affecting decision to cross, is crucial to formulate evidence-based measures to enhance safe mobility and remove barriers to walking. Numerous studies have been conducted over the past two decades to dissect the pedestrian gap-acceptance behaviors in front of human-driven vehicles, as summarized in Table A1 in Appendix A.

Built on the questionnaire survey and controlled experiment, earlier studies have investigated the gap-acceptance by instructing participants to indicate their crossing intentions in hypothetical scenarios via textual descriptions (Holland and Hill, 2007; Zhou et al., 2009). While such a research design facilitates the incorporation of unobservable psychological features such as personal traits, habits, subjective norms, attitudes, and risk perceptions, its ecological validity raises doubt, because of the failure to couple perception with action (Lobjois et al., 2013). Liu and Tung (2014) then invited 32 participants to watch the pre-recorded, real-world videos in which a human-driven motor vehicle was approaching an unprotected, two-lane crosswalk. By pressing the space bar on the keyboard to indicate road-crossing intention, the elderly pedestrians were found to make the same crossing decision as the younger, without realizing their lower safety margins arisen from the decline in walking abilities. To further improve realism, Soathong et al. (2021) conducted an onsite questionnaire survey with 400 participants at four unprotected mid-block road sections without any crossing facilities and developed the SEM to untangle the motivation behind risky crossing behaviors. Their study highlights that pedestrians' intention to cross was predominantly driven by their habits and attitude.

With the rapid progress of sensor and computer-vision techniques, another common practice is the onsite video observation, which captures the real-life pedestrian crossing behaviors using high-definition videos at distinct locations, such as the midblock crosswalks (Cherry et al., 2012; Liu and Tung, 2014; Kadali et al., 2015; Pawar and Patil, 2015; 2016; Naser et al., 2017; Zhang et al., 2018; Avinash et al., 2019; Zhang et al., 2019; Zhao et al., 2019; Kadali and Vedagiri, 2020; Alver et al., 2021; Pawar and Yadav, 2022; Angulo et al., 2023), signalized intersections (Koh and Wong, 2014; Onelcin and Alver, 2015; Zafri, 2023), unsignalized intersections (Vasudevan et al., 2020; Zafri et al., 2020; Zafri, 2023), multi-lane road sections (Zhuang and Wu, 2011; Shaaban et al., 2018; Sheykhfard and Haghighi, 2020; Zhang et al., 2024), and overpasses (Demiroz et al., 2015; Alver and Onelcin, 2018). Gap-acceptance has been found to depend on a variety of contextual factors, including the pedestrian age, gender, trip purpose, luggage carrying, group size, waiting time, crossing speed, crossing strategy (single-stage, two-stage, or rolling gap), vehicle approaching-speed, time to arrival, vehicle size, vehicle type, the number of traffic lanes, the presence of pedestrian facilities, weather, and lighting conditions. Specifically, elderly pedestrians tended to accept a larger time gap (Koh and Wong, 2014; Petzoldt, 2014; Kadali et al., 2015; Naser et al., 2017), while pedestrians waiting at the median (Zhao et al., 2019), the adoption of rolling gap crossing behaviors (Kadali et al., 2015; Naser et al., 2017; Zhang et al., 2018; Zafri et al., 2020), longer waiting time (Kadali et al., 2015; Lobjois et al., 2013), larger pedestrian group sizes (Kadali et al., 2015), more vehicles passed (Lobjois et al., 2013), approaching of slower and lighter vehicles (Sheykhfard and Haghighi, 2020;

Alver et al., 2021), more attempts made by pedestrians to cross (Kadali et al., 2015), and the presence of a physical barrier (Alver et al., 2021) were associated with a smaller gap accepted. Based on the logistic regression model, the probability of gap acceptance was found to decrease with smaller gap sizes (Cherry et al., 2012; Naser et al., 2017; Zhao et al., 2019; Kadali and Vedagiri, 2020; Sheykhfard and Haghighi, 2020; Vasudevan et al., 2020; Pawar and Yadav, 2022), shorter waiting time (Cherry et al., 2012; Zhao et al., 2019; Sheykhfard and Haghighi, 2020; Vasudevan et al., 2020), wider crosswalks (Zhao et al., 2019; Kadali and Vedagiri, 2020; Vasudevan et al., 2020), higher vehicle-approaching speeds (Cherry et al., 2012; Petzoldt, 2014; Pawar and Patil, 2015; Naser et al., 2017; Kadali and Vedagiri, 2020; Sheykhfard and Haghighi, 2020; Pawar and Yadav, 2022), and the approaching of larger vehicles (Pawar and Patil, 2015; Naser et al., 2017; Kadali and Vedagiri, 2020; Pawar and Yadav, 2022). Arguably, in reality, pedestrian crossing decision is not solely determined by the individual demographic and situational characteristics. Although substantial effort has been made to complement video recording with onsite questionnaire surveys to encompass motivational factors (Arellana et al., 2022), such experimental paradigm remains confined to existing traffic scenarios, incapable of exploring pedestrian interactions with unprecedented technologies such as automated truck platoons.

Recent advancements in VR technology have enabled the simulation of diversified traffic scenarios in an interactive, controllable, reproducible, and risk-free environment (Schneider and Bengler, 2020; Tran et al., 2021; Li et al., 2025a). Compared with simply watching the pre-recorded videos in front of computer screens or projectors (Lobjois and Cavallo, 2007; 2009; Liu and Tung, 2014; Petzoldt, 2014; Madigan et al., 2019; Dey et al., 2019; Rad et al., 2020; Leung et al., 2021; Zhao et al., 2022; Shen et al., 2023; Feng et al., 2024), VR allows participants to experience more immersive, realistic interactions by 3D near-eye displays and pose tracking. Woodman et al. (2019) conducted a VR experiment with head-mounted displays (HMD) to investigate the gap-acceptance of a platoon with four automated vehicles, approaching at a constant speed (i.e., 1 km/h, 4 km/h, 8 km/h, or 16 km/h) and a fixed time gap (i.e., 2 s, 3 s, 4 s, or 5 s) on shared space without road markings and pavements or on a single-lane road with pavement and grass verge. Interestingly, pedestrians were found more likely to accept a smaller time gap in the shared environment, although lower perceived safety was reported. Despite being informative, their study hardly draws definitive conclusions given a limited sample size of 28 participants. Tian et al. (2022) developed a visual looming model to explain the pedestrian gap-acceptance at single-lane, unsignalized midblock crosswalks. After analyzing the crossing behaviors of 60 participants in front of a platoon of two automated vehicles operating at a constant approaching-speed (i.e., 25 mph, 30 mph, or 35 mph) and a preset time gap (i.e., 2 s, 3 s, 4 s, or 5 s) in the cave automatic virtual environment (CAVE), they reported that pedestrians tended to accept a smaller gap when vehicles approached at a higher speed and the visual looming could largely explain such speed-induced unsafe crossing behavior. Although the CAVE has been proven to be highly immersive, self-explanatory, and user-friendly (Dommes et al., 2014; Sobhani and Farooq, 2018; Tapiro et al., 2018; O'Neal et al., 2019; Tapiro et al., 2020; Ye et al., 2020; Dommes et al., 2021; Jiang et al., 2021; Soares et al., 2021; Lee et al., 2022; Tian et al., 2022; 2023; Kalantari et al., 2023; Ye et al., 2024), the large-scale screens and rear projectors are indeed costly and space-consuming (Li et al., 2025a). Recently, the HMD has become increasingly popular due to its portable and cost-effective merits (Deb et al., 2018; Sobhani and Farooq, 2018; Morrongiello et al., 2019; Nuñez Velasco et al., 2019; Woodman, et al., 2019; Deb et al., 2020; Bindschädel et al., 2021; Camara et al., 2021; Kalatian and Farooq., 2021; Nuñez Velasco et al., 2021; Kwon et al., 2022; Wang et al., 2022; Angulo et al., 2023; Figueroa-Medina et al., 2023; Song et al., 2023; Bennett et al., 2025; Feng et al., 2025). Figueroa-Medina et al. (2023) conducted an HMD-based VR experiment with 48 subjects to simulate pedestrian crossing behaviors on unprotected midblock crosswalks. Despite that versatile behavioral data such as the number of gaps watched before crossing, gap accepted to cross, time waited to cross, walking speed, and crossing success rate have been deliberately collected, the absence of fine-grained psychological indicators restricts our ability to comprehend the internal mechanisms underlying behavioral tendency, risk perception, and decision-making.

In summary, previous studies have obtained valuable insights into how pedestrians behave when interacting with human-driven or automated vehicles. These findings, however, may not be readily generalizable to pedestrian gap-acceptance of automated truck platoons. On one hand, pedestrians likely adjust their behaviors by waiting longer time, accepting a larger time gap or crossing faster when confronted with a truck, because of its uniquely huge size and reduced maneuverability. For another, the barrier effect caused by organizing several trucks in a platoon, together with the relatively slower operation speed, may prompt pedestrians to engage in risk-taking behaviors. The motivation underlying such decision-making process, however, remain largely unknown. To fill this gap, by coupling the self-reported questionnaire survey with the HMD-based VR simulator, the present study focuses particularly on how pedestrians' behavioral tendency, trust, and risk perception shape their crossing intentions in front of automated truck platoons.

**2.3 Behavioral tendency, trust, and risk perception in crossing decision**

Although onsite observations and VR simulation experiments can capture versatile, subtle behavioral cues, it seems impracticable to encompass all types of pedestrian behaviors under dynamic, unconstrained situations. To measure the tendency to perform risky behaviors in a cost-effective, safe, and reproducible manner, Granié et al. (2013) introduced the Pedestrian Behavior Scale, with 47 survey items stratified into a four-factor structure, i.e., transgressions, lapses, aggressive behaviors, and positive behaviors. By adapting some items to local settings, Qu et al. (2016) developed the Chinese version of Pedestrian Behavior Scale. The lack of sufficient internal consistency in the lapse subscale, together with the inadequacy of measuring aggressive behaviors (Qu et al., 2016; Xu et al., 2018; Shen et al., 2023), however, questions its validity to Chinese population (Vandroux et al., 2022). Soon afterwards, Deb et al. (2017a) refined the original Pedestrian Behavior Scale and proposed the PBQ, which categorized pedestrian behaviors as violations, errors, lapses, aggressive behaviors, and positive behaviors. Four items with the highest factor loadings from each of the five subscales were then compiled to form a short version. Such a 20-item version of PBQ has been validated across multiple countries, including the US (Deb et al., 2017a; 2017b; 2018), UK (McIlroy et al., 2019; 2020), China (McIlroy et al., 2019; 2020), Thailand (McIlroy et al., 2019; 2020), Vietnam (McIlroy et al., 2019; 2020; Dinh et al., 2020), Bangladesh (McIlroy et al., 2019; 2020), and Kenya (McIlroy et al., 2019; 2020). Specifically, by leveraging a 7-item demographic questionnaire, a 4-item personal innovativeness scale, the short version of PBQ, and a 16-item questionnaire of pedestrian receptivity toward AVs, Deb et al. (2017b) conducted an online survey with 482 participants in the US to evaluate their behavioral intentions to cross the road with AVs. The results indicated that pedestrians who complied with traffic rules and showed positive behaviors toward other road users believed that the promotion of AVs would improve overall traffic safety, while those with higher scores in violations, lapses, and aggressive behaviors felt more confidence to cross in front of AVs. Subsequently, after gauging both objective and subjective measures from 30 participants in an HMD-based VR experiment in Starkville, US, Deb et al. (2018) found that pedestrians who frequently committed errors or showed aggressive behaviors toward other road users waited less before stepping into the crosswalk, took longer time to cross in front of AVs, and rated poorly for AVs even with external features. By contrast, pedestrians who often violated traffic rules intentionally were more cautious and appreciated the inclusion of external interfaces in AVs. Rad et al. (2020), however, reached an opposite conclusion that pedestrians who reported more violation behaviors were less likely to yield to AVs.

In addition to the behavioral tendency, trust in AVs also plays a nonnegligible role in affecting pedestrian crossing decisions. Zhao et al. (2022) explored the intention to cross in front of AVs or human-driven vehicles in risky situations. Based on an online questionnaire survey collected from 493 respondents in Australia, their study found that participants held greater trust, perceived lower risks, and held more positive attitudes toward road-crossing in front of AVs than human-driven vehicles. Similar findings have also been reported by Rad et al. (2020) and Feng et al. (2024) that pedestrians with a higher level of trust in AVs showed a stronger willingness to cross. Nuñez Velasco et al. (2019) conducted a VR experiment with 55 participants to investigate the intention to cross the road in front of AVs with different physical appearance and external interfaces in Delft, Netherlands. Surprisingly, although pedestrians who recognized the vehicle as automated reported a higher level of trust, they generally exhibited a lower intention to cross when an AV approached. With the rapid development of automated technology and gradual penetration of AVs, trust in AVs may evolve accordingly (Wu et al., 2023). It is therefore essential to further discern whether the varying degrees of trust that pedestrians have in AVs will ultimately affect their gap-acceptance behaviors.

As Table A2 in Appendix B presents, as a subjective judgment on potential traffic hazards, risk perception also influences pedestrian behaviors in the presence of AVs. Kown et al. (2022) analyzed how the risks perceived by pedestrians affected their crossing behaviors at a four-way, unsignalized intersection in a residential block. Based on a VR experiment with 200 young participants in Ulsan and Seoul, South Korea, their study indicated that pedestrians with higher perceived risks took longer time to start walking and tended to cross in haste. By a virtual video-based online questionnaire survey with 589 participants in China, Feng et al. (2024) reached a similar conclusion that those with a higher level of perceived risks were more conservative regarding road-crossing in front of AVs.

To summarize, previous studies typically investigated the individual contribution of behavioral tendency, trust, and risk perception to pedestrian crossing decisions, without taking their interplay into account. Understanding such an interdependent relationship helps not only explain what pedestrians behave, but also reflect how they perceive, decide, and adapt when interacting with AVs.

### 2.4 SEM-ANN model

Many important attributes such as trust and risk perception cannot be observed directly. As a powerful data-analytic technique, SEM has been widely used in social, behavioral, and health sciences, given its flexibility in representing latent constructs and estimating the relationships between observed and unobserved variables simultaneously (Zhou et al., 2009; Soathong et al., 2021; Ye et al., 2021; Kwon et al., 2022). Soathong et al. (2021) developed the SEM to specify the casual relationship between motivational factors (i.e., attitudes, social norms, perceived behavioral controls, and habits) and pedestrian intention to cross at midblock road sections. Likewise, Kwon et al. (2022) employed the hierarchical SEM to investigate the environment–perception–behavior interaction and found that the perceived risk mediated the environmental influences on pedestrian crossing behaviors. However, SEM is only able to model the linear relationships, which oversimplifies the inherent complexities of pedestrian crossing decisions (Deng et al., 2018).

To address this limitation, we introduce a novel hybrid, two-stage approach by leveraging the strengths of both SEM and ANN. The SEM is adept at establishing theoretical models and testing causal-effect hypotheses, while the ANN outperforms in capturing nonlinear relationships and uncovering interconnected patterns. Although recent studies have demonstrated the SEM-ANN as a versatile tool in predicting fashion shopping behaviors (Ng et al., 2022), gauging consumer interests in virtual gifts of esports (Chen and Wu, 2024), evaluating student readiness for robotics in education (Suhail et al., 2024), and predicting cryptocurrency adoptions (Arpaci and Bahari, 2023), such a promising method has not received sufficient attention among traffic psychologists and travel behavior analysts. By integrating the hypothesis-testing capability of SEM with the nonlinear modeling merit of ANN, we present a new avenue to address the interconnection between motivations and behavioral intentions.

## 3. Methods

### 3.1 Research framework

As Fig. 1 illustrates, our study comprises four components, i.e., hypothesis making, experimental design, questionnaire survey, and SEM-ANN analysis. First, a hypothetical model is developed to integrate key indicators, including technological understanding, trust in safety, behavioral tendency, risk perception, and gap-acceptance behavior, along with their interrelationships.

This setup establishes the theoretical foundation for subsequent experiment designs.

Second, the VR technology is harnessed to create first-person perspective videos that simulate realistic scenarios involving automated truck platoons. These controlled scenarios allow participants to make crossing decisions, ensuring ecological validity.

Third, the VR-based questionnaire aligned with the hypothetical model is designed as the primary data collection instrument to systematically capture participants' both subjective measures and objective behavioral responses to simulated scenarios.

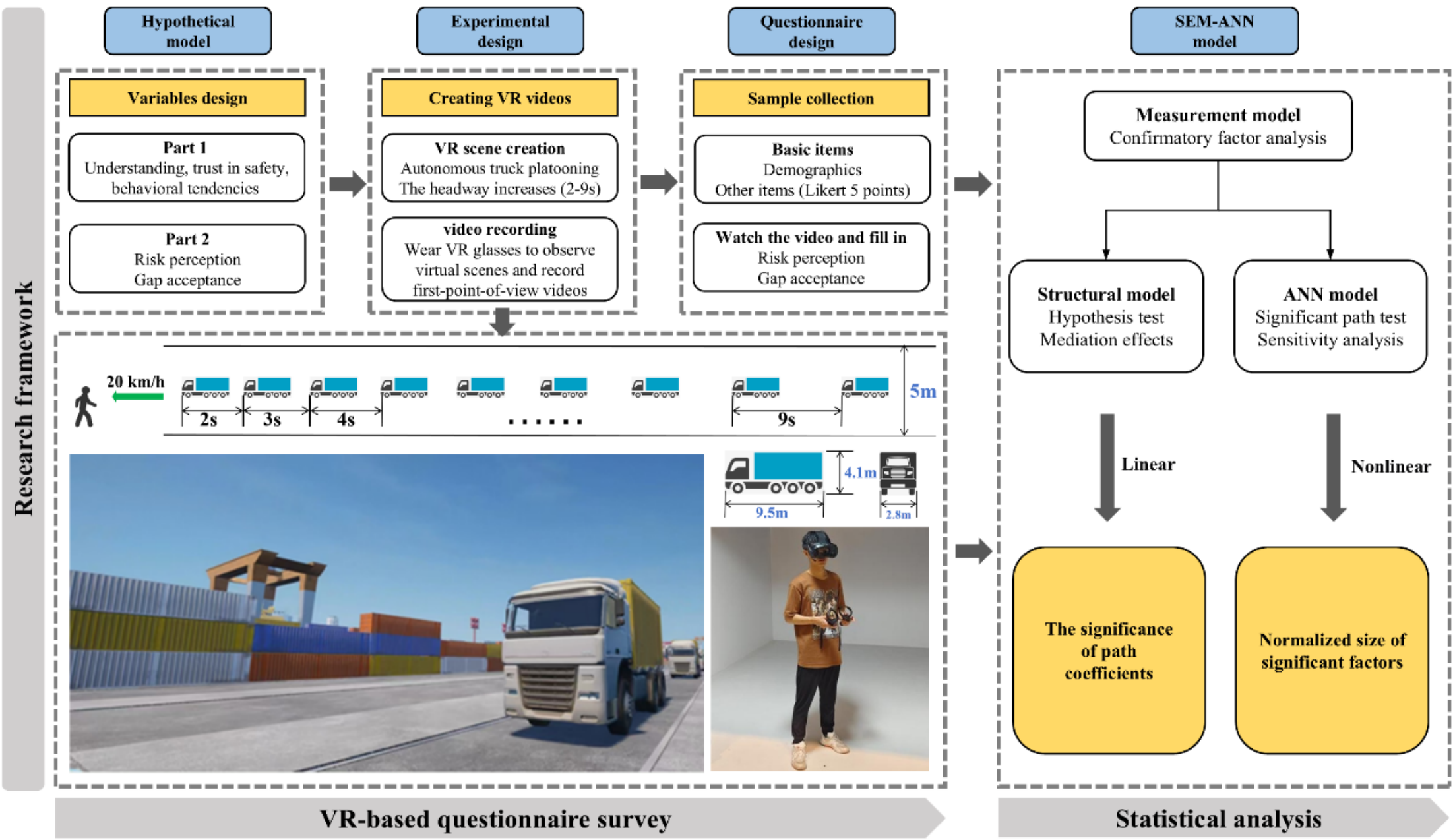

**Fig. 1.** An overview of the research framework.

Last, the hybrid SEM-ANN model is developed. The SEM first tests the hypothesized relationships among variables, which help uncover the causal pathways and validate the theoretical framework. Then, the ANN captures the nonlinear interactions between significant predictors and outcomes, revealing nuanced behavioral patterns.

### 3.2 Hypothetical model

As Table A3 in Appendix C presents, three types of latent variables were specified to characterize motivational factors underlying pedestrian crossing decisions. First, the well-validated short version of PBQ (Deb et al., 2017a) was adopted to quantify the tendency of behaviors performed by pedestrians in daily life. We then stratified trust-related constructs into five dimensions as understanding of conventional trucks, understanding of automated driving technology, trust in the safety of conventional trucks, trust in the safety of automated driving technology, and trust in the safety of automated trucks. Last, we evaluated the risk perception of crossing in front of automated truck platoons following Kwon et al. (2022) and Feng et al. (2024). All the items were measured by a five-point Likert scale, ranging from one (strongly disagree) to five (strongly agree).

According to the technology acceptance theory, understanding of an emerging technology influences an individual's trust, attitude, and receptivity (Lourenço et al., 2024). As knowledge accumulates, people foster stronger expectations that the system will act safely and predictably, thereby raising trust. Based on this rationale, the following hypotheses are made:

**H1:** Understanding of conventional trucks increases pedestrian trust in the safety of trucks.

**H2:** Understanding of automated driving technology increases pedestrian trust in the safety of automated driving technology.

**H3:** Understanding of conventional trucks enhances pedestrian trust in the safety of automated trucks.

**H4:** Understanding of automated driving technology enhances pedestrian trust in the safety of automated trucks.

The trust transfer theory posits that trust in one entity can be transferred to another related target entity (Lee and Hong, 2019). If pedestrians possess greater trust in the safety of conventional trucks, such trust potentially extends to automated trucks due to similarities in vehicle appearance and function performance. Accordingly, we have:

**H5:** Trust in the safety of conventional trucks positively affects trust in the safety of automated trucks.

**H6:** Trust in the safety of automated driving technology positively affects trust in the safety of automated trucks.

Given that pedestrians with a higher level of trust in AVs and a lower level of risk perception showed a stronger willingness to cross in front of AVs (Zhao et al., 2022; Feng et al., 2024), we make the following hypotheses:

**H7:** Greater trust in the safety of automated trucks is associated with the acceptance of smaller time gaps.

**H8:** A higher level of risk perception is associated with the acceptance of larger time gaps.

Likewise, over-trust in the safety of automated trucks likely lowers pedestrians' own risk perception during road-crossing. Thereby, we have:

**H9:** Greater trust in the safety of automated trucks decreases the level of risk perception.

In addition, pedestrians who comply with traffic rules and show positive behaviors toward other road users might be risk-averse:

**H10:** Behavioral tendency toward positive behaviors is associated with a higher level of risk perception.

Last, based on the findings that pedestrians who often committed errors or showed aggressive behaviors waited less before stepping into the crosswalks (Deb et al., 2018) and those who frequently violated traffic rules intentionally were less likely to yield to AVs (Rad et al., 2020), we propose the following hypothesis:

**H11:** Negative behavioral tendencies such as violations, errors, and aggressive behaviors are associated with the acceptance of smaller gaps.

Fig. 2 illustrates the proposed theoretical framework.

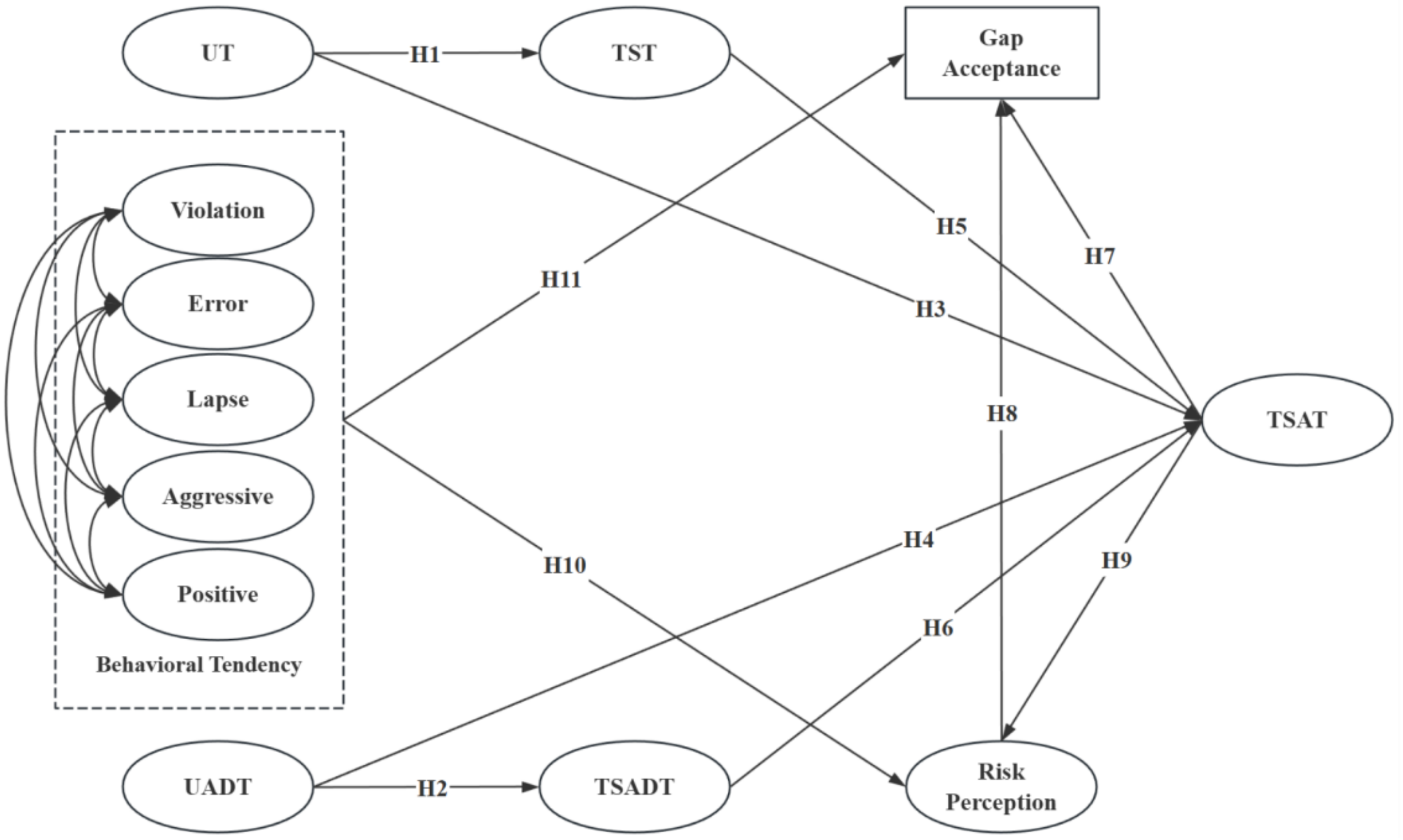

**Fig. 2.** The proposed theoretical framework (UT: Understanding of trucks; UADT: Understanding of automated driving technology; TST: Trust in the safety of trucks; TSADT: Trust in the safety of automated driving technology; TSAT: Trust in the safety of automated trucks).

### 3.3 Data collection

In the present study, the VR video-based questionnaire survey was conducted to collect data on pedestrian behavioral tendency, trust, risk perception, and gap-acceptance of automated truck platoons, given its advantages in internal validity, reproducibility, and cost-effectiveness.

**Experiment scenario design:** The participants experienced a virtual traffic situation from the viewpoint of pedestrians. The scenario was created by the HMD. A platoon consisted of nine automated trucks, each with 9.5 m in length, 2.8 m in width, and 4.1 m in height. The trucks were approaching from the right-side of the participant at a constant speed of 20 km/h, with headways increasing incrementally from 2 s to 9 s. The participant was standing at the curbside of a single-lane road on a sunny afternoon. This type of interaction is representative in the suburb of China, where pedestrians most likely encounter truck platoons in the near future as on-road testing continues. Consistent with Rodríguez Palmeiro et al. (2018), de Clercq et al. (2019), Dey et al. (2019), Dommes et al. (2021), Lee et al. (2022), and Feng et al. (2025), we chose not to use the zebra crossing, because in China the trucks are obliged to yield when a pedestrian stands on the curb and is about to cross a marked crosswalk.

**Experiment procedure:** As Fig. 3 shows, we adopted the VR video-based, questionnaire-embedded approach implemented via an online survey platform. First, the consent form was signed and participants were instructed to fill a questionnaire to delineate their background information, behavioral tendency, understanding of trucks, trust in the safety of trucks, understanding of automated technology, trust in the safety of automated driving technology, and trust in the safety of automated trucks. Afterwards, the participants were presented with a first-person-perspective video generated by the VR, where a truck platoon was approaching from 50 m away. The participants were not allowed to make crossing decisions before the first truck arrived and were instructed to press the button at the last moment that they felt safe to cross (de Clercq et al., 2019; Feng et al., 2025). Then, participants were invited to fill a questionnaire to measure their perceived risk of road-crossing. It took about 15 min to complete the whole experiment.

To ensure data integrity, three quality-control measures were implemented:

- ◆ **Time control:** Minimum 80 s were required for each assessment module.
- ◆ **Device limit:** Users with the same IP address could only submit once.
- ◆ **Response verification:** logically inconsistent entries were automatically excluded (e.g., the claim of holding a driving license by the under-age).

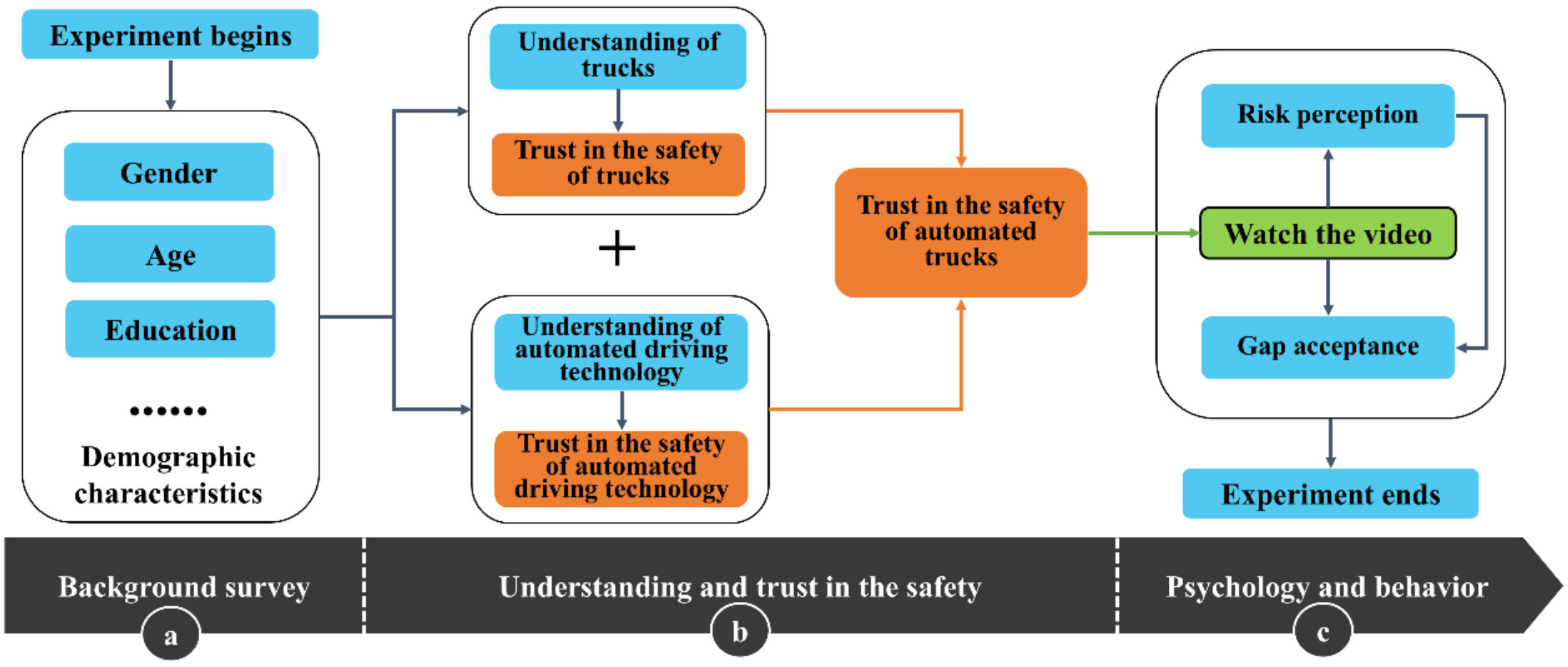

**Fig. 3.** Process of the VR video-based questionnaire survey.

**Participant recruitment:** Using personal invitations, website registrations, and campus posters, we attracted 603 individuals to participant in our survey, whose basic demographic characteristics are presented in Table 1. Note that the subjects were stratified into two subgroups by whether they were younger than 35 years or not, given that the number of old participants was relatively scarce and further subdividing age into finer intervals would reduce the statistical significance of model estimations. In China, the age of 35 also represents a socially and institutionally accepted threshold to distinguish younger populations in employment practices and public discourse.

Ethical approval was obtained from the Human Research Ethics Committee of Ningbo University.

**Table 1.** Descriptive statistics of demographic characteristics of participants.

| **Variable** | **Frequency** | **Proportion** |
|---|---|---|
| Gender | | |
| Female | 148 | 24.54% |
| Male | 455 | 75.46% |
| Age | | |
| $< 35$ | 477 | 79.10% |
| $\geqslant 35$ | 126 | 20.90% |
| Education level | | |
| Junior high school or below | 8 | 1.33% |
| High school | 114 | 18.91% |
| College student | 400 | 66.34% |
| Postgraduate student or above | 81 | 13.43% |
| Driving license | | |
| Yes | 566 | 93.86% |
| No | 37 | 6.14% |
| Years of driving experience | | |
| 0 | 49 | 8.13% |
| $< 1$ | 115 | 19.07% |
| 1–3 | 233 | 38.64% |
| 3–5 | 108 | 17.91% |
| $> 5$ | 98 | 16.25% |
| Professional experience in the field of transportation | | |
| Yes | 241 | 39.97% |
| No | 362 | 60.03% |
| Collision experience | | |
| Yes | 330 | 54.73% |
| No | 273 | 45.27% |

**Note:** Collision experience refers to whether participants had previously been involved in traffic collisions as pedestrians, cyclists, or other vulnerable road users.

### 3.4 SEM-ANN model specification

To elicit the interdependence among psychological constructs and observed gap-acceptance decisions, a two-stage, hybrid approach combining SEM and ANN was developed. The SEM was first specified to examine causal relationships and test hypotheses under linear assumptions, followed by ANN applied to capture the potential nonlinear associations between significant predictors and outcome variables identified by SEM.

**(1) Stage 1: SEM formulation**

As a powerful tool that combines factor analysis and multiple regression analysis to investigate relationships among multiple variables, SEM encompasses two main components: the measurement model and the structural model. The former specifies the relationship between observed variables and their corresponding latent variables, while the latter establishes the relationship between latent constructs. Specifically, the measurement model is formulated as:

$$x_i = \Lambda_x \eta_i + \delta_i \tag{1}$$

$$\mathrm{y}_i = \Lambda_y \xi_i + \varepsilon_i \tag{2}$$

where $\eta_i$ and $\xi_i$ represent the $i$th unobserved endogenous and exogenous variables, respectively. $x_i$ and $\mathrm{y}_i$ are vectors of the observed exogenous and endogenous indicators, respectively. $\Lambda_x$ and $\Lambda_y$ specify how the observed indicators are linked to the unobserved constructs. Here, $\varepsilon_i$ and $\delta_i$ denote the error terms of the indicators.

Meanwhile, the structural model is specified as:

$$\eta_j = A\eta_i + B\xi_i + v_i \tag{3}$$

where $A$ and $B$ are the matrices of structural parameters to be estimated. $v_i$ denotes the vector of unobserved errors.

To calibrate the SEM, the confirmatory factor analysis (CFA) was first performed to verify the relationships between observed indicators and underlying constructs. Once the reliability and validity of the measurement model were confirmed, the maximum likelihood estimation was used to quantify the relationship between latent constructs. Finally, bootstrapping, a non-parametric statistical model, was employed to estimate the significance and magnitude of mediating effects. By repeatedly sampling the original dataset, bootstrapping allows to calculate the distribution and confidence intervals of mediating effects, thereby shedding light on the role of trust in mediating the relationship between technological understanding and risk perception, and the role of risk perception in mediating the relationship between behavioral tendency and gap-acceptance.

### (2) Stage 2: ANN formulation

As a computational model that simulates the behavior of brain neurons, the ANN can map complex relationship between input and output data. Specifically, the input layer receives raw data, the hidden layers are responsible for extracting features, while the output layer produces predictive results. Each layer comprises multiple neurons connected by weights. Signals are propagated forward through the network and transformed by activation functions. The bias parameter is added to adjust for the output of a neuron. Unlike weights, bias parameters are not tied to any specific input but instead shift the activation function to better fit the data. Thereby, biases are internal calibration and are not treated as predictors in the sensitivity analysis.

Fig. 4 illustrates the pipeline of ANN model designed to simulate relationships among latent variables in the SEM. The input layer includes significant exogenous and endogenous variables estimated from the SEM, which are transmitted through weighted connections to two hidden layers with nodes and bias terms. After processing through the hidden layers, the endogenous variable of the SEM is predicted in the output layer.

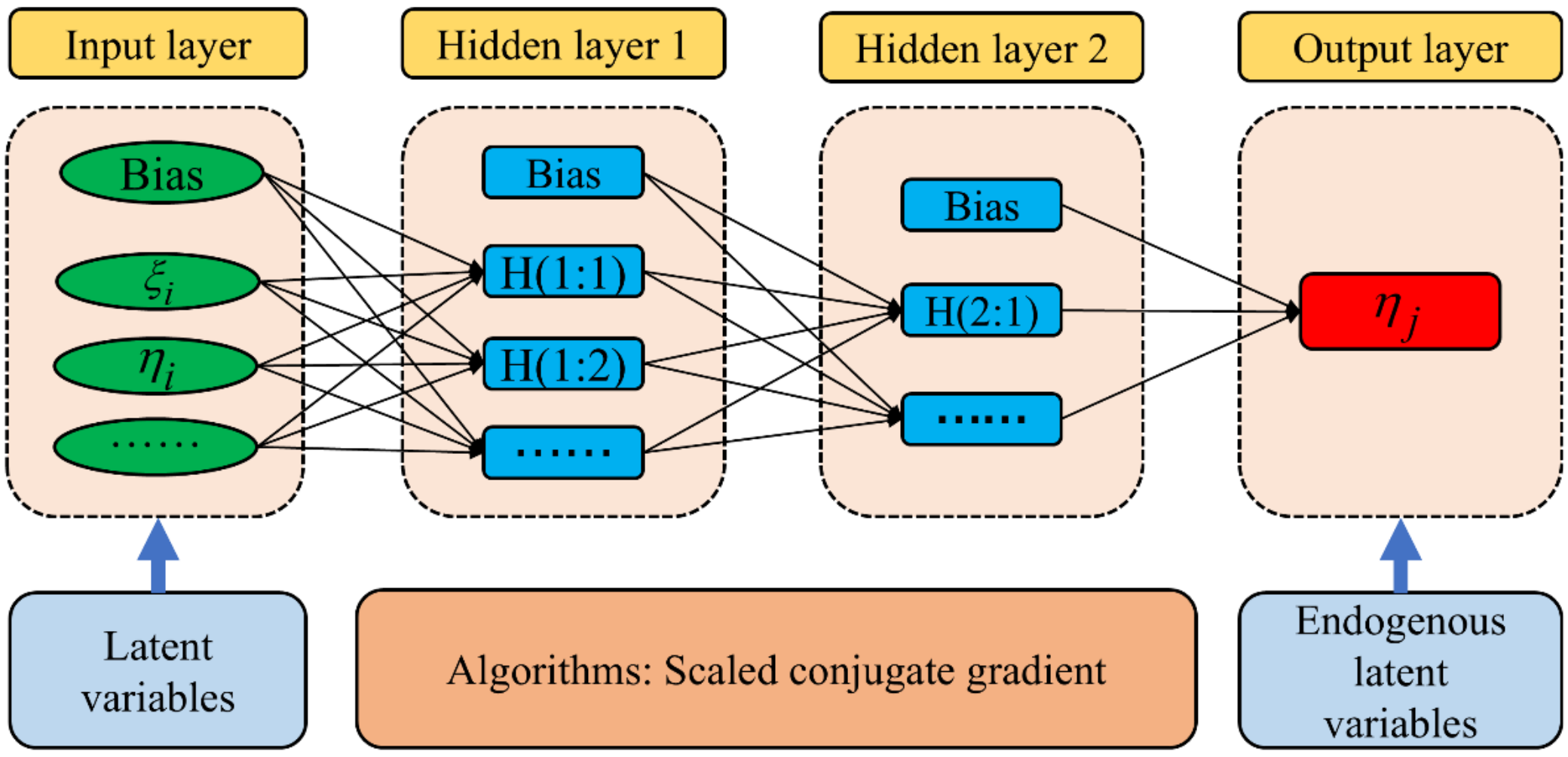

**Fig. 4.** The ANN part of the hybrid SEM-ANN approach.

As a key component of ANN, activation functions determine whether the output of a neuron should be activated. The sigmoid function was chosen as the activation function for both hidden and output neurons, because of its capacity in smoothing outputs and introducing nonlinearity into models. To alleviate the risk of overfitting, the 10-fold cross-validation test was performed, in which the original observations were randomly partitioned into 10 mutually exclusive subsets of equal size. Each time, one subset was left out as the validation while the remaining was combined for model estimation. The scale conjecture gradient method was used to fine-tune variable weights, with the root mean square error (RMSE) as the evaluation metric.

# 4. Results

## 4.1 Descriptive analysis

As Table A4 in Appendix D presents, on average, participants watched about five vehicle gaps before deciding to cross and accepted a gap of 5.35 s. Only 0.82% of participants accepted a gap of 2 s, and 10.11% accepted a gap equal to or less than 3 s.

The one-way analysis of variance (ANOVA) was used to assess the influence of demographic factors on target variables of interest. Insignificant associations at the 95% confidence level were not included for parsimony. As Table 2 and Fig. 5 illustrate, older participants reported a deeper understanding of conventional trucks and held greater trust in the safety of automated trucks. Those with driving license showed a better understanding of conventional trucks and automated driving technology, and had greater trust in the safety of automated driving technology and automated trucks. Participants who had previously involved in traffic collisions also reported a better understanding of conventional trucks and held greater trust in the safety of trucks.

**Table 2.** Results of one-way ANOVA (***: $p < 0.001$; **: $p < 0.01$; *: $p < 0.05$).

| **Independent variable** | **Dependent variable** | **Degrees of freedom** | $F$ | $\eta_p^2$ |
|---|---|---|---|---|
| Age | Understanding of trucks | (1,601) | 15.88*** | 0.02 |
| Driving license | | (1,601) | 25.45*** | 0.04 |
| Collision experience | | (1,601) | 17.19*** | 0.03 |
| Education level | | (3,599) | 10.53*** | 0.05 |
| Driving experience | | (4,598) | 25.45*** | 0.08 |
| Transport practitioner | | (1,601) | 34.40*** | 0.05 |
| Collision experience | Trust in the safety of trucks | (1,601) | 3.98* | 0.01 |
| Transport practitioner | | (1,601) | 10.74** | 0.02 |
| Driving license | Understanding of automated driving technology | (1,601) | 24.88*** | 0.00 |
| Education level | | (3,599) | 3.74** | 0.02 |
| Driving experience | | (4,598) | 24.88*** | 0.06 |
| Transport practitioner | | (1,601) | 25.09*** | 0.04 |
| Driving license | Trust in the safety of automated driving technology | (1,601) | 7.32** | 0.01 |
| Education level | | (3,599) | 8.21*** | 0.04 |
| Driving experience | | (4,598) | 7.32*** | 0.03 |
| Transport practitioner | | (1,601) | 9.72** | 0.02 |
| Age | Trust in the safety of automated trucks | (1,601) | 4.07* | 0.04 |
| Driving license | | (1,601) | 6.49* | 0.01 |
| Education level | | (3,599) | 5.00*** | 0.02 |
| Driving experience | | (4,598) | 6.49*** | 0.04 |
| Transport practitioner | | (1,601) | 13.16*** | 0.02 |
| Transport practitioner | Risk perception | (1,601) | 5.18* | 0.01 |

In addition, education level, driving experience, and transport practitioner were significantly associated with understanding of trucks, understanding of automated driving technology, trust in the safety of automated driving technology, and trust in the safety of automated trucks. Interestingly, only transport practitioners perceived higher risks for road-crossing in front of truck platoons. These findings indicated that trust, risk perception, and gap acceptance varied substantially across subjects. Thereby, demographic factors were included as controls in the SEM to mitigate confounding effects.

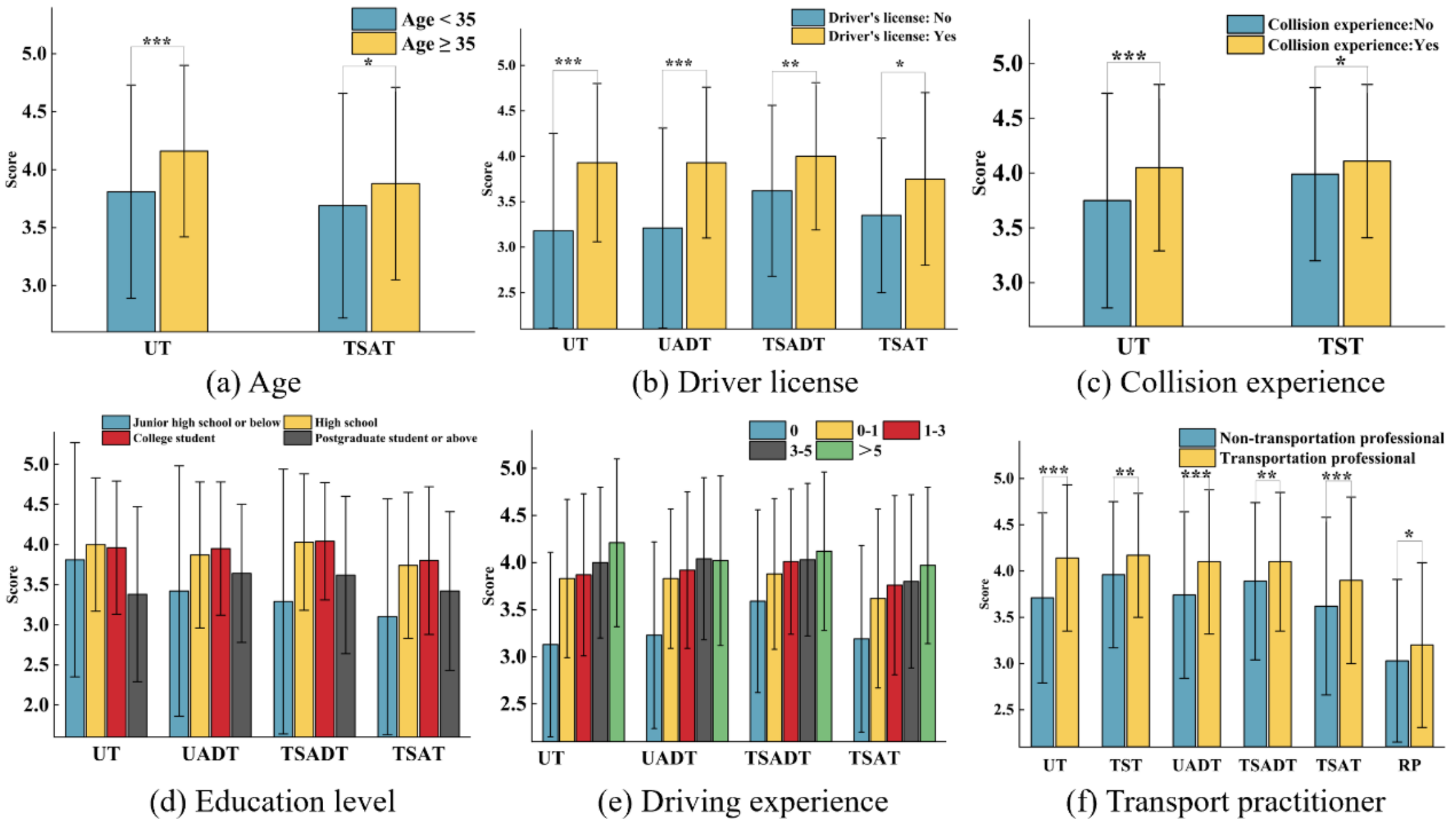

**Fig. 5.** The impact of demographics on target variables (***: $p < 0.001$; **: $p < 0.01$; *: $p < 0.05$; UT: understanding of trucks; TST: trust in the safety of trucks; UADT: understanding of automated driving technology; TSADT: trust in the safety of automated driving technology; TSAT: trust in the safety of automated trucks; RP: risk perception).

### 4.2 Measurement model

The CFA was used to evaluate the structural validity of measurement model. The results are presented in Table 3.

**Table 3.** Goodness-of-fit indices of the measurement model (RMSEA: root-mean-square error of approximation; GFI: goodness-of-fit index; CFI: comparative fit index; TLI: Tucker–Lewis index; $df$ : degrees of freedom).

| Fit category | Fit index | Value | Criterion of acceptance |
|---|---|---|---|
| Absolute fit | RMSEA | 0.04 | < 0.08 (Wheaton, 1987) |
| | GFI | 0.90 | > 0.90 (Miles and Shevlin, 2007) |
| Incremental fit | CFI | 0.96 | > 0.90 (Miles and Shevlin, 2007) |
| | TLI | 0.96 | > 0.90 (Miles and Shevlin, 2007) |
| Parsimonious fit | $\chi^2$ / $df$ | 1.75 | < 5.00 (Marsh and Hocevar, 1985) |

As Table 3 shows, all the fit metrics met or exceeded the recommended thresholds, suggesting that the measurement model achieves a satisfactory overall fit. Detailed CFA model with standardized estimates is presented in Appendix E.

Besides goodness-of-fit measures, the factor loadings, Cronbach's alpha, average variance extracted, and composite reliability were calculated to assess the validity and reliability of latent constructs. As Table 4 presents, all the items were highly significant, with $p$-value less than 0.001. The Cronbach's alpha values were greater than 0.70 and the average variance extracted values were above 0.50, confirming sufficient reliability and convergent validity (Awang, 2012). The square root of average variance extracted also exceeded the correlations between constructs, as shown in Table A5 in Appendix F. Altogether, these results demonstrate that the developed measurement model is reliable, supporting the validity of subsequent structural model analysis.

**Table 4.** Results of the factor loading, Cronbach's alpha, composite reliability, and average variance extracted for measurement model (UT: understanding of trucks; TST: trust in the safety of trucks; UADT: understanding of automated driving technology; TSADT: trust in the safety of automated driving technology; TSAT: trust in the safety of automated trucks; RP: risk perception).

| | **Estimate** | **Cronbach's alpha** | **Average variance extracted** | **Composite reliability** |
|---|---|---|---|---|
| UT → UT1 | 0.74 | 0.86 | 0.61 | 0.86 |
| UT → UT2 | 0.76 | | | |
| UT → UT3 | 0.80 | | | |
| UT → UT4 | 0.82 | | | |
| TST → TST1 | 0.76 | 0.80 | 0.59 | 0.81 |
| TST → TST2 | 0.70 | | | |
| TST → TST3 | 0.83 | | | |
| UADT → UADT1 | 0.78 | 0.84 | 0.62 | 0.83 |
| UADT → UADT2 | 0.79 | | | |
| UADT → UADT3 | 0.78 | | | |
| TSADT → TSADT1 | 0.81 | 0.82 | 0.60 | 0.82 |
| TSADT → TSADT2 | 0.77 | | | |
| TSADT → TSADT3 | 0.75 | | | |
| Violation → Violation1 | 0.81 | 0.87 | 0.63 | 0.87 |
| Violation → Violation2 | 0.77 | | | |
| Violation → Violation3 | 0.82 | | | |
| Violation → Violation4 | 0.791 | | | |
| Error → Error1 | 0.76 | 0.84 | 0.57 | 0.84 |
| Error → Error2 | 0.78 | | | |
| Error → Error3 | 0.68 | | | |
| Error → Error4 | 0.78 | | | |
| Lapse → Lapse1 | 0.76 | 0.89 | 0.67 | 0.89 |
| Lapse → Lapse2 | 0.85 | | | |
| Lapse → Lapse3 | 0.80 | | | |
| Lapse → Lapse4 | 0.86 | | | |
| Aggressive → Aggressive1 | 0.85 | 0.91 | 0.71 | 0.91 |
| Aggressive → Aggressive2 | 0.84 | | | |
| Aggressive → Aggressive3 | 0.86 | | | |
| Aggressive → Aggressive4 | 0.83 | | | |
| Positive → Positive1 | 0.75 | 0.87 | 0.60 | 0.86 |
| Positive → Positive2 | 0.78 | | | |
| Positive → Positive3 | 0.79 | | | |
| Positive → Positive4 | 0.78 | | | |
| TSAT → TSAT1 | 0.83 | 0.91 | 0.63 | 0.91 |
| TSAT → TSAT2 | 0.80 | | | |
| TSAT → TSAT3 | 0.83 | | | |
| TSAT → TSAT4 | 0.75 | | | |
| TSAT → TSAT5 | 0.74 | | | |
| TSAT → TSAT6 | 0.80 | | | |
| RP → RP1 | 0.71 | 0.81 | 0.52 | 0.81 |
| RP → RP2 | 0.78 | | | |
| RP → RP3 | 0.67 | | | |
| RP → RP4 | 0.71 | | | |

### 4.3 Structural model

Based on the results of CFA, the structural model was calibrated using the maximum likelihood estimation. As Table 5 presents, only the goodness-of-fit index for the absolute fit failed marginally to reach the recommended threshold. Given that in multivariate models it seems challenging to satisfy all the fit criteria (Schreiber et al., 2010), our structural model is broadly acceptable.

**Table 5.** Goodness-of-fit indices of the structural model (RMSEA: root-mean-square error of approximation; GFI: goodness-of-fit index; CFI: comparative fit index; TLI: Tucker–Lewis index; $df$ : degrees of freedom).

| Fit category | Fit index | Value | Criterion of acceptance |
|---|---|---|---|
| Absolute fit | RMSEA | 0.04 | $< 0.08$ (Wheaton, 1987) |
| | GFI | 0.88 | $> 0.90$ (Miles and Shevlin, 2007) |
| Incremental fit | CFI | 0.94 | $> 0.90$ (Miles and Shevlin, 2007) |
| | TLI | 0.93 | $> 0.90$ (Miles and Shevlin, 2007) |
| Parsimonious fit | $\chi^2 / df$ | 1.95 | $< 5.00$ (Marsh and Hocevar, 1985) |

Table 6 summarizes the results of hypothesis testing, and Fig. 6 illustrates the simplified path diagram with standardized estimates. Note that only significant relationships between latent constructs were included here.

**Table 6.** Results of hypothesis testing (***: $p < 0.001$; **: $p < 0.01$; *: $p < 0.05$).

| Hypothesis | Description | Estimate | Result |
|---|---|---|---|
| H1 | Understanding of trucks → Trust in the safety of trucks | 0.48*** | True |
| H2 | Understanding of automated driving → Trust in the safety of automated driving | 0.71*** | True |
| H3 | Understanding of trucks → Trust in the safety of automated trucks | 0.28*** | True |
| H4 | Understanding of automated driving → Trust in the safety of automated trucks | 0.37*** | True |
| H5 | Trust in the safety of trucks → Trust in the safety of automated trucks | 0.14*** | True |
| H6 | Trust in the safety of automated driving → Trust in the safety of automated trucks | 0.38*** | True |
| H7 | Trust in the safety of automated trucks → Gap-acceptance | –0.02 | False |
| H8 | Risk perception → Gap-acceptance | 0.49*** | True |
| H9 | Trust in the safety of automated trucks → Risk perception | –0.17* | True |
| H10 | Positive behaviors → Risk perception | 0.23** | True |
| H11 | Positive behaviors → Gap-acceptance | 0.30*** | True |
| | Violations → Gap-acceptance | –0.16** | True |

Specifically, understanding of trucks was positively associated with the trust in the safety of trucks ($\beta = 0.48$, $p < 0.001$) and the trust in the safety of automated trucks ($\beta = 0.28$, $p < 0.001$), supporting hypothesis H1 and H3, respectively. Similarly, understanding of automated driving technology positively influenced the trust in the safety of automated driving technology ($\beta = 0.71$, $p < 0.001$) and the trust in the safety of automated trucks ($\beta = 0.37$, $p < 0.001$), supporting hypothesis H2 and H4, respectively. Likewise, hypotheses H5 and H6 were also tenable, given that the paths from trust in the safety of trucks and from trust in the safety of automated driving technology to the trust in the safety of automated trucks were statistically significant at the 99% confidence interval with an estimate of 0.14 and 0.38, respectively. Surprisingly, the trust in the safety of automated trucks had no direct effects on gap-acceptance ($\beta = -0.02$, $p > 0.05$), but yielded a significantly negative association with risk perception ($\beta = -0.17$, $p < 0.05$). Thereby, hypothesis H7 was rejected, while H9 held true.

Furthermore, there was a significantly positive relationship between risk perception and gap-acceptance ($\beta = 0.49$, $p < 0.001$), supporting hypothesis H8 that pedestrians with a higher level of risk perception tended to accept a larger time gap. In terms of behavioral tendency, positive behaviors were associated with a higher level of risk perception ($\beta = 0.23$, $p < 0.01$) and acceptance of larger time gaps ($\beta = 0.30$, $p < 0.001$), while errors and aggressive behaviors were negatively associated with risk perception. Notably, violations directly affected gap acceptance ($\beta = -0.16$, $p < 0.05$), supporting hypothesis H11 that pedestrians who often violated traffic rules were more likely to accept a smaller time gap.

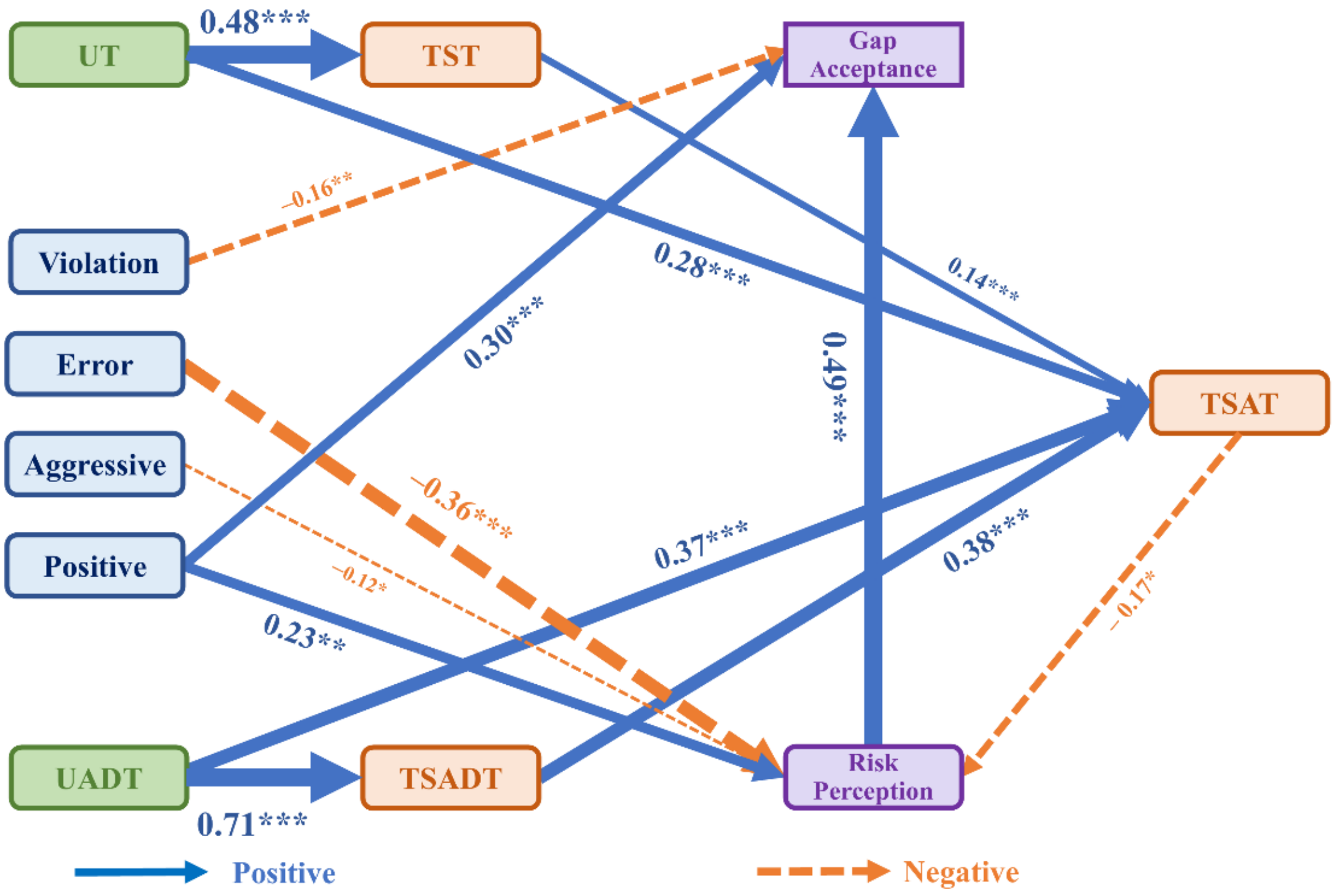

**Fig. 6.** Diagram of SEM with standardized estimates of significant paths (***: $p$ <0.001; **: $p$ <0.01; *: $p$ <0.05; UT: understanding of trucks; TST: trust in the safety of trucks; UADT: understanding of automated driving technology; TSADT: trust in the safety of automated driving technology; TSAT: trust in the safety of automated trucks).

Table 7 presents the effects of demographic characteristics on latent variables. Specifically, understanding of trucks increased with age ($\beta = 0.11$, $p < 0.01$), driving experience ($\beta = 0.18$, $p < 0.001$), and transport-related profession ($\beta = 0.23$, $p < 0.01$), but decreased with education level ($\beta = -0.15$, $p < 0.01$). By contrast, trust in the safety of trucks was higher among the well-educated ($\beta = 0.09$, $p < 0.05$), but lower for the older ($\beta = -0.09$, $p < 0.05$) and the female ($\beta = -0.09$, $p < 0.05$). In addition, understanding of automated driving technology was deeper among experienced drivers ($\beta = 0.16$, $p < 0.01$) and transport practitioners ($\beta = 0.22$, $p < 0.01$), while licensed drivers reported lower trust in the safety of automated trucks ($\beta = -0.10$, $p < 0.01$).

**Table 7.** The effects of demographic characteristics on latent variables.

| Path | Estimate | *p*-value |
|---|---|---|
| Age → Understanding of trucks | 0.11 | <0.01 |
| Age → Trust in the safety of trucks | –0.09 | <0.05 |
| Driving experience → Understanding of trucks | 0.18 | <0.001 |
| Driving experience → Understanding of automated driving | 0.16 | <0.01 |
| Driving license → Trust in the safety of automated trucks | –0.10 | <0.01 |
| Education level → Understanding of trucks | –0.15 | <0.01 |
| Education level → Trust in the safety of trucks | 0.09 | <0.05 |
| Gender → Trust in the safety of trucks | –0.09 | <0.05 |
| Transport practitioner → Understanding of trucks | 0.23 | <0.01 |
| Transport practitioner → Understanding of automated driving | 0.22 | <0.01 |

### 4.4 Mediation effects

The mediating effects were analyzed via bootstrapping, with a particular focus on the interplay among trust in the safety of automated trucks, risk perception, and gap-acceptance. As Table 8 shows, in addition to the direct influence, understanding of trucks also indirectly affected the trust in the safety of automated trucks through trust in the safety of trucks. Similar results could also be observed that the influence of understanding of automated driving technology on the trust in the safety of automated trucks was partially mediated by the trust in the safety of automated driving technology.

**Table 8.** Mediation effects (***: $p < 0.001$; **: $p < 0.01$; *: $p < 0.05$; UT: understanding of trucks; TST: trust in the safety of trucks; UADT: understanding of automated driving technology; TSADT: trust in the safety of automated driving technology; TSAT: trust in the safety of automated trucks; RP: risk perception).

| Relation | Direct effect | Indirect effect | Mediation |
|---|---|---|---|
| UT → TST → TSAT | 0.28*** | 0.07** | Partial |
| UADT → TSADT → TSAT | 0.37** | 0.27*** | Partial |
| TST → TSAT → RP | –0.07 | –0.02* | Full |
| TSADT → TSAT → RP | –0.03 | –0.06* | Full |
| TSAT → RP → Gap-acceptance | –0.03 | –0.08* | Full |
| Positive → RP → Gap-acceptance | 0.30*** | 0.12** | Partial |
| Aggressive → RP → Gap-acceptance | 0.06 | –0.06* | Full |
| Errors → RP → Gap-acceptance | 0.02 | –0.17** | Full |

Notably, the effects of trust in the safety of trucks and trust in the safety of automated driving technology on risk perception were fully mediated by the trust in the safety of automated trucks. These results indicate that trust in the safety of automated trucks plays a crucial mediating role in shaping risk perception. Also importantly, risk perception serves as the bond between pedestrian behavior tendency and crossing decision, as the effects of trust in the safety of automated trucks, aggressive behaviors, and errors on gap-acceptance were fully mediated by risk perception. Risk perception also partially mediated the relationship between positive behaviors and gap-acceptance.

### 4.5 ANN Analysis

Given that SEM could only examine linear relationships, ANN was developed to untangle nonlinear relationships. Based on the significant variables identified by SEM, separate models were developed for trust in the safety of automated trucks, risk perception, and gap-acceptance. To evaluate the prediction accuracy, the RMSE for both the training and testing phases of ANN was calculated. As Table 9 presents, the RSME ranged from 0.008 to 0.022, indicating that the ANN achieves a favorable fit to the original data and can effectively generalize to new, unseen data. The relatively low standardized deviation further demonstrates the consistency and stability of model performance.

**Table 9.** RMSE for ANN Model A, B and C, with the trust in the safety of automated trucks, risk perception, and gap-acceptance as the output variable, respectively (SD: Standard deviation. RMSE: root mean square error).

| | **Model A** | | **Model B** | | **Model C** | |
|---|---|---|---|---|---|---|
| Cross-validation | Training RMSE | Testing RMSE | Training RMSE | Testing RMSE | Training RMSE | Testing RMSE |
| Round 1 | 0.013 | 0.009 | 0.018 | 0.017 | 0.011 | 0.012 |
| Round 2 | 0.012 | 0.010 | 0.018 | 0.014 | 0.013 | 0.009 |
| Round 3 | 0.011 | 0.012 | 0.018 | 0.016 | 0.013 | 0.011 |
| Round 4 | 0.011 | 0.011 | 0.017 | 0.016 | 0.012 | 0.009 |
| Round 5 | 0.011 | 0.011 | 0.017 | 0.018 | 0.012 | 0.011 |
| Round 6 | 0.011 | 0.013 | 0.017 | 0.012 | 0.011 | 0.009 |
| Round 7 | 0.012 | 0.008 | 0.017 | 0.014 | 0.012 | 0.011 |
| Round 8 | 0.011 | 0.011 | 0.017 | 0.018 | 0.012 | 0.015 |
| Round 9 | 0.011 | 0.012 | 0.018 | 0.016 | 0.011 | 0.010 |
| Round 10 | 0.012 | 0.012 | 0.017 | 0.022 | 0.012 | 0.014 |
| Mean | 0.011 | 0.011 | 0.017 | 0.016 | 0.012 | 0.011 |
| SD | 0.001 | 0.001 | 0.001 | 0.003 | 0.001 | 0.002 |

To quantify the contribution of each feature to the output variable, the permutation feature importance was calculated by randomly shuffling the values of a single feature and observing the resulting degradation of model performance (Pedregosa et al., 2011). We then averaged the importance score across the 10-fold cross-validation and computed the relative importance by normalizing its value against that of the most influential input variable.

As Table 10 presents, understanding of automated driving technology had the greatest influence on trust in the safety of automated trucks (100%), followed by trust in the safety of automated driving technology (71.39%), understanding of trucks (65.44%), trust in the safety of trucks (37.39%), and driving license status (8.78%).

**Table 10.** Importance of input variables on trust in the safety of automated trucks (UT: understanding of trucks; TST: trust in the safety of trucks; UADT: understanding of automated driving technology; TSADT: trust in the safety of automated driving technology; NRI: normalized relative importance).

| Cross-validation | UT | TST | UADT | TSADT | Driving license |
|---|---|---|---|---|---|
| Round 1 | 0.27 | 0.12 | 0.30 | 0.30 | 0.01 |
| Round 2 | 0.20 | 0.19 | 0.30 | 0.27 | 0.04 |
| Round 3 | 0.19 | 0.14 | 0.42 | 0.23 | 0.02 |
| Round 4 | 0.22 | 0.16 | 0.36 | 0.22 | 0.04 |
| Round 5 | 0.24 | 0.08 | 0.36 | 0.25 | 0.07 |
| Round 6 | 0.22 | 0.14 | 0.35 | 0.26 | 0.02 |
| Round 7 | 0.23 | 0.12 | 0.38 | 0.24 | 0.03 |
| Round 8 | 0.22 | 0.11 | 0.36 | 0.28 | 0.03 |
| Round 9 | 0.25 | 0.09 | 0.37 | 0.26 | 0.04 |
| Round 10 | 0.26 | 0.17 | 0.35 | 0.21 | 0.02 |
| Mean | 0.23 | 0.13 | 0.35 | 0.25 | 0.03 |
| NRI | 65.44% | 37.39% | 100.00% | 71.39% | 8.78% |

As Table 11 presents, tendency to error behaviors had the greatest influence on risk perception (100%), followed by positive behaviors (64.45%), trust in the safety of automated trucks (58.96%), and aggressive behaviors (51.73%).

**Table 11.** Importance of input variables on risk perception (TSAT: trust in the safety of automated trucks; NRI: normalized relative importance).

| Cross-validation | Errors | Aggressive | Positive | TSAT |
|---|---|---|---|---|
| Round 1 | 0.31 | 0.17 | 0.27 | 0.19 |
| Round 2 | 0.37 | 0.17 | 0.21 | 0.20 |
| Round 3 | 0.30 | 0.19 | 0.28 | 0.19 |
| Round 4 | 0.33 | 0.19 | 0.23 | 0.22 |
| Round 5 | 0.35 | 0.18 | 0.20 | 0.22 |
| Round 6 | 0.36 | 0.17 | 0.17 | 0.24 |
| Round 7 | 0.38 | 0.16 | 0.15 | 0.21 |
| Round 8 | 0.36 | 0.16 | 0.25 | 0.21 |
| Round 9 | 0.35 | 0.22 | 0.26 | 0.15 |
| Round 10 | 0.36 | 0.18 | 0.21 | 0.20 |
| Mean | 0.35 | 0.18 | 0.22 | 0.20 |
| NRI | 100.00% | 51.73% | 64.45% | 58.96% |

As Table 12 presents, risk perception had the greatest influence on gap-acceptance (100%), followed by tendency to violation (43.50%) and positive (50.87%) behaviors.

**Table 12.** Importance of input variables on gap-acceptance (NRI: normalized relative importance).

| Cross-validation | Violations | Positive | Risk perception |
|---|---|---|---|
| Round 1 | 0.19 | 0.25 | 0.57 |
| Round 2 | 0.28 | 0.26 | 0.45 |
| Round 3 | 0.32 | 0.27 | 0.42 |
| Round 4 | 0.19 | 0.26 | 0.55 |
| Round 5 | 0.20 | 0.33 | 0.47 |
| Round 6 | 0.24 | 0.24 | 0.52 |
| Round 7 | 0.20 | 0.29 | 0.51 |
| Round 8 | 0.18 | 0.25 | 0.58 |
| Round 9 | 0.18 | 0.21 | 0.62 |
| Round 10 | 0.26 | 0.27 | 0.46 |
| Mean | 0.22 | 0.26 | 0.52 |
| NRI | 43.50% | 50.87% | 100.00% |

We ultimately compared the path coefficients derived from SEM and the normalized relative importance scores obtained from ANN. As Table 13 presents, the estimation results of SEM and ANN were broadly consistent. Solely the importance rankings of understanding of automated driving technology and trust in the safety of automated driving technology on trust in the safety of automated trucks differed slightly.

**Table 13.** Comparison between SEM and ANN results (TSAT: trust in the safety of automated trucks; UT: understanding of trucks; TST: trust in the safety of trucks; UADT: understanding of automated driving technology; TSADT: trust in the safety of automated driving technology).

| Input variables | SEM path coefficient | ANN normalized relative importance | SEM ranking | ANN ranking | Match |
|---|---|---|---|---|---|
| **Output variable: TSAT** | | | | | |
| UT | 0.28 | 65.44% | 3 | 3 | Yes |
| TST | 0.14 | 37.39% | 4 | 4 | Yes |
| UADT | 0.37 | 100.00% | 2 | 1 | No |
| TSADT | 0.38 | 71.39% | 1 | 2 | No |
| Driving license | –0.10 | 8.78% | 5 | 5 | Yes |
| **Output variable: risk perception** | | | | | |
| Errors | –0.36 | 100.00% | 1 | 1 | Yes |
| Aggressive behaviors | –0.12 | 51.73% | 4 | 4 | Yes |
| Positive behaviors | 0.23 | 64.45% | 2 | 2 | Yes |
| TSAT | –0.17 | 58.96% | 3 | 3 | Yes |
| **Output variable: gap-acceptance** | | | | | |
| Risk perception | 0.49 | 100.00% | 1 | 1 | Yes |
| Positive behaviors | 0.30 | 50.87% | 2 | 2 | Yes |
| Violations | –0.16 | 43.50% | 3 | 3 | Yes |

## 5. Discussion

In this section, we first analyze the time gap accepted by participants briefly. We then uncover the role of motivational factors in shaping pedestrian crossing decisions in front of automated truck platoons, with a particular focus on the interplay among behavioral tendency, trust, risk perception, and gap-acceptance. The new insights derived from the SEM-ANN are also highlighted. A range of evidence-based countermeasures are finally proposed to ensure safer and smoother interactions between pedestrians and automated truck platoons.

### 5.1 Time gap accepted

In our experiment, participants had to analyze the approaching speed of truck platoons and the available gaps between trucks before making the decision to cross the one-lane street. The results indicated that subjects watched an average of five gaps between trucks before deciding to cross and the average time gap accepted was about 5.35 s (with a standard deviation of 1.43), which was substantially larger than that (4.00 s with a standard deviation of 1.03) reported by Figueroa-Medina et al. (2023), who conducted a VR experiment with 48 subjects to investigate pedestrian behavioral cues when crossing a single-lane, marked midblock crosswalk in front of AV platoons. This finding is largely expected, because the unique characteristics of trucks, e.g., large size, heavy weight, and limited maneuverability, likely pose greater visual pressure to pedestrians. Previous studies also found that pedestrians tended to accept a larger time gap when larger vehicles were approaching (Sheykhfard and Haghighi, 2020; Alver et al., 2021).

### 5.2 Influence of behavioral tendency and risk perception

Previous studies have consistently proven that behavioral tendency significantly influences both observable behaviors and psychological responses (Vandroux et al., 2022). Ignoring such heterogeneity can result in oversimplified models that fail to capture the complexity of real-world behaviors. By leveraging the well-validated PBQ (Deb et al., 2017a), we successfully characterized the tendency of pedestrians to perform various behaviors in daily life and demonstrate that there exists complex interplay among behavioral tendency, risk perception, and gap-acceptance of automated truck platoons. Consistent with Rad et al. (2020) that pedestrians who reported more violation behaviors tended to cross before the approaching AV stopped completely, our study also found that subjects who violated traffic rules intentionally were more likely to accept a smaller time gap. By contrast, those who complied with traffic rules and showed positive behaviors to other road users were more inclined to wait and accept a time gap larger by approximately 30%. This result is expected largely, because positive behavioral tendency is often associated with higher safety awareness and stronger adherence to traffic norms, which collectively reinforce cautious decision-making. Based on a VR experiment with 30 participants in Starkville, US, Deb et al. (2018) reported a similar finding that pedestrians with a positive behavioral tendency spent more time to observe and understand the external features of AVs before starting crossing.

As the subjective judgment that individuals make regarding the characteristics and consequence of risks, our study demonstrates that risk perception plays the most dominant role in how pedestrians respond to potential hazards. By simultaneously including behavioral tendency, risk perception, and trust into the gap-acceptance model, we provide new insights that in addition to the direct association, risk perception also serves as the strong bond, which fully mediated the effects of aggressive behaviors, errors, and trust in the safety of automated trucks on gap-acceptance. Kwon et al. (2022) also indicated that risk perception fully mediated the relationship between environmental attributes and pedestrian crossing behaviors at a four-way, unsignalized intersection, based on a VR experiment with 200 young participants in South Korea. Indeed, individuals likely adjust their behaviors to changes in perceived risks. It is thus unsurprising that subjects with a higher level of perceived risks tended to accept a larger time gap. Similar results were also reported by Kwon et al. (2022) and Feng et al. (2024). Another finding worthy of mention is that tendency to errors and aggressive behaviors was negatively associated with risk perception. That is, pedestrians who frequently committed errors and showed aggressive behaviors to other road users reported a lower level of risk for road-crossing in front of automated truck platoons. Errors are usually results of lack of knowledge or misunderstandings of traffic rules and aggressiveness is a personality. Therefore, the risk of interactions with automated trucks is likely to be underestimated by those who do not have enough knowledge about traffic systems and those who always expect to have the right of way.

### 5.3 Influence of trust-related constructs

In addition to the behavioral tendency and risk perception, trust in AVs, known as individual belief that an AV will perform its intended task with high effectiveness (Deb et al., 2017b), potentially shapes pedestrian crossing decisions when interacting with AVs. To delve into the role of trust in gap-acceptance of automated truck platoons, we characterized trust as understanding of conventional trucks and automated driving technology, trust in the safety of conventional trucks and automated driving technology, and trust in the safety of automated trucks. Although Rad et al. (2020), Zhao et al. (2022), and Feng et al. (2024) found that a higher level of trust in AVs increased pedestrian willingness to cross in front of AVs, our results indicated that trust-related constructs had no direct associations with gap-acceptance. Instead, the effects of trust in the safety of automated trucks on gap-acceptance were fully mediated by risk perception. That is, pedestrians with a higher level of trust in the safety of automated trucks tended to accept a smaller time gap because of a lower risk perceived. Such indirect effects of trust on acceptance of various technologies and hazards (e.g., gene technology and nuclear power) have also been addressed by Siegrist (2021).

Furthermore, when assessing the safety of automated trucks, pedestrians likely draw their prior knowledge about conventional trucks and automated technology. It is thus unsurprising that understanding of conventional trucks and automated technology was positively associated with trust in the safety of automated trucks. Similarly, given that trust in one entity can be transferred to related target entity (Lee and Hong, 2019), participants who held greater trust in conventional trucks and automated technology were found to show greater trust in automated trucks. Another interesting finding is that understanding of automated driving technology yielded the greatest impact on trust in the safety of automated trucks. Thereby, prompting the public to better understand AV's ability to navigate safely seems the most effective and straightforward means to build trust in automated trucks.

### 5.4 Influence of demographic factors

The demographic characteristics of participants were included to address individual heterogeneity, and we found significant influences of age, gender, education level, driving experience, driver's license status, and transport-related professions on understanding and trust constructs. Specifically, younger and female subjects reported greater trust in the safety of conventional trucks, whereas the older exhibited greater trust in the safety of automated trucks. One plausible explanation is that individuals with limited exposure to truck operations may overestimate the safety of conventional trucks, whereas more experienced pedestrians likely recognize the potential of automated technology to enhance the safety performance of trucks, thereby exhibiting higher trust in the safety of automated trucks (Navarro et al., 2025). Likewise, extensive driving experience and professional experience in transportation were associated with a better understanding of both conventional trucks and automated driving technologies, while participants with a valid driving license showed lower trust in the safety of automated trucks, a finding align with Holland and Hill (2007) that licensed drivers were more sensitive to the risk of crossing in front of AVs than non-drivers.

Unlike previous studies that reported a direct relationship between demographic characteristics and crossing decisions (Holland and Hill, 2007; Salducco et al., 2022; Feng et al., 2024), our study found that the effects of demographic factors on risk perception and gap-acceptance of automated truck platoons were not significant, but fully mediated by trust in the safety of automated trucks. Based on an onsite observation with 444 jaywalkers at six two-lane signalized intersections, Onelcin and Alver (2015) indicated that pedestrian age and gender did not result in a significant association with gap-acceptance behaviors. Zhao et al. (2022) also revealed that the effects of age and gender on pedestrian intention to cross in front of AVs became negligible once motivational factors such as attitude and trust were included.

### 5.5 New insights from SEM-ANN

Although SEM offers clear statistical evidence of causal and mediating relationships, it overlooks the potential nonlinear relationships between variables. By contrast, while the ANN outperforms in capturing interdependence, the black box nature limits its applications for hypothesis testing. In the present study, we proposed the SEM-ANN, a hybrid, two-stage model which leverages the strengths of both SEM and ANN. Although a comparison of the path coefficients derived from SEM and the relative importance scores obtained from ANN indicated that the estimation results of SEM and ANN were broadly consistent, the effects of trust in the safety of automated driving technology on the trust in the safety of automated trucks were indeed underestimated by the SEM, potentially due to the failure of SEM in addressing the nonlinearity. In this regard, the SEM-ANN provides a promising tool for traffic psychologists and ergonomists to model intricate relationships between latent variables.

### 5.6 Implications

Based on the aforementioned findings, tailor-made measures are proposed following the “3E” (engineering, education, and enforcement) principle.

- **Engineering:** Given the dominant role of risk perception on crossing decisions, policymakers should prioritize refining communication mechanisms between pedestrians and automated trucks. For example, incorporating non-verbal signals, such as light cues or auditory warnings into vehicle communication systems, can help compensate for the lack of eye contact and enhance the trust in automated technology (Lee and Hess, 2020; Rezwana and Lownes, 2024). By explicitly conveying the vehicle's intention (e.g., yielding or not), such external human-machine interfaces are expected to mitigate uncertainty, increase confidence, and reduce risk perception of pedestrians when interacting with automated truck platoons (Rouchitsas and Alm, 2019; Fabricius et al., 2022). In addition, since frequent decelerations undermine the benefits of truck platoons and 99.18 % of participants were found to reject a time gap equal to or less than 2 s, maintaining truck gaps short enough to prevent pedestrians from cutting in is promoted on arterial roads. To reduce potential conflicts between pedestrians and automated trucks in highly populated areas, another effective countermeasure is to segregate these vulnerable road users from truck platoons by setting dedicated crossing facilities such as overpasses and underpasses.

- **Education:** In terms of education, our study suggests that enhancing understanding of automated trucks is crucial. Traffic management authorities, vehicle manufactures, and local communities should work collaboratively to address public concerns and develop best practices. Through systematic education initiatives such as publicity, courses, and virtual reality training to elucidate the operation principles and safety features of heavy trucks and automated driving technology, local authorities can dispel misconceptions, alleviate fears, and improve trust in automated truck platoons (Lourenço et al., 2024; Feng et al., 2024). Specific educational programs should also be promoted to improve the safety awareness of pedestrians who often commit errors and show aggressive behaviors toward other road users, as these individuals were found more inclined to underestimate risks and make reckless crossings.
- **Enforcement:** Our findings also have profound implications for traffic legislation and enforcement. With the gradual penetration of AVs, pedestrians likely adjust their crossing behaviors. Policymakers should balance pedestrian safety with traffic efficiency when introducing automated truck platoon into public roads. Intelligent reminder systems such as sound signals and visual cues can be mandatorily deployed to alert pedestrians of the approaching and intentions of automated trucks. Given that violation behavioral tendency resulted in a significant and direct effect on risk-taking gap-acceptance behaviors, regulations against jaywalking in front of automated trucks should be strictly enforced.

## 6. Conclusions

This study attempted to address the gap in understanding motivations underlying pedestrian intention to cross the road in front of automated truck platoons by exploring the relationships among behavioral tendency, trust, risk perception, and gap-acceptance. The VR video-based questionnaire survey was conducted to collect both latent psychological measures and observed gap-acceptance behaviors. The SEM-ANN model was developed to examine both linear and nonlinear relationships among latent variables.

For the first time, we shed lights on how pedestrians perceive and respond to automated truck platoons in road-crossing tasks. The results indicate that pedestrian gap-acceptance of automated truck platoons was predominantly influenced by risk perception and behavioral tendency. Specifically, subjects with a higher level of risk perception and showing positive behaviors to other road users were more likely to accept a larger time gap, while those who frequently violated traffic rules tended to accept a smaller time gap. Also importantly, risk perception served as the connection among behavioral tendency, trust, and gap-acceptance. Those who showed positive behaviors to other road users generally reported a higher level of risk perception, while tendency to error and violation behaviors and over-trust in the safety of automated trucks were associated with a lower level of risk perception. In addition, participants who had a better understanding and more trust in conventional trucks and automated driving technology also showed more trust in the safety of automated trucks.

Given these findings, evidence-based countermeasures were proposed in terms of engineering, education, and enforcement to address the unique challenges posed by automated truck platoons, which help foster safer and more cooperative interactions between pedestrians and automated systems.

Our study, however, is not without limitations. First, the trucks were approaching at a constant speed of 20 km/h, with headways increasing incrementally from 2 s to 9 s. Although such scenario design helps enhance internal validity, it potentially impairs external validity. Future studies should embrace continuous and real-time interactions, where both human participants and automated trucks can perceive and respond to each other's actions. Second, given that we focused solely on pedestrian intention to cross a single-lane, unmarked midblock road section on a sunny day, our findings might not be readily generalized to marked midblock crossings, multi-lane roads, signalized intersections, rainy days, and nighttime. Third, since the elderly are hardly reached by online questionnaires (Feng et al., 2024), our samples were skewed toward young subjects. Stratified sampling should be employed in future work to achieve a more balanced age cohort. Last, to enhance the ecological validity of VR results, reconstructing more realistic scenes from the real-world images (Angulo et al., 2023) and designing virtual experiments based on the real-life datasets (Bennett et al., 2025) are highly encouraged.

**Declaration of competing interest**

The authors declare that they have no known competing financial interests or personal relationships that could have appeared to influence the work reported in this paper.

**Appendix A**

**Table A1.** A summary of representative studies on pedestrian gap-acceptance behaviors in front of human-driven vehicles over the past two decades. Studies not explicitly modeling the gaps accepted by pedestrians (e.g., pedestrian preferences towards different crossing facilities) are not included here.

| Study | Region | Experiment | Scenario | Target variable | Analysis method | Major findings |
|---|---|---|---|---|---|---|
| Holland and Hill (2007) | Birmingham, UK | Questionnaire survey with 293 participants | Three hypothetical scenarios with high risks | Intention to cross the road | Theory of planned behavior and structural equation model | ◆ The female and elderly perceived more risk and were less likely to cross in risky situations. |
| Lobjois and Cavallo (2007) | Caen, France | Virtual video-based survey with 78 participants | A single-lane urban street | Time/distance gap accepted, crossing time, response time, and safety margin | Logistic regression model | ◆ Elderly participants accepted a larger time gap for crossing than the younger.<br>◆ Given a time constraint, all age groups accepted a smaller time gap at a higher vehicle-approaching speed.<br>◆ Without time constraint, the younger selected similar time gaps regardless of vehicle speed, whereas the elderly accepted a smaller time gap at a higher vehicle-approaching speed. |
| Lobjois and Cavallo (2009) | Caen, France | Virtual video-based survey with 78 participants | A single-lane urban street | Time gap accepted, initiation crossing time, crossing time, and safety margin | Logistic and linear regression models | ◆ Elderly participants accepted a larger time gap and initiated crossing sooner.<br>◆ The elderly accepted a smaller time gap when vehicle approached faster.<br>◆ All groups adjusted their crossing time to the available time. |
| Zhou et al. (2009) | Beijing, China | Questionnaire survey with 426 participants | Two hypothetical scenarios which violated the traffic signal to cross the street | Crossing intention | Theory of planned behavior and structural equation model | ◆ The elderly had a lower intention to cross.<br>◆ Respondents reported a greater intention to cross when other pedestrians were crossing the road.<br>◆ Respondents with high conformity tendency had stronger intentions to cross.<br>◆ Intention to cross decreased, as perceived risk increased, attitudes and subjective norms became more negative, and perceived behavioral control decreased. |
| Zhuang and Wu (2011) | Hangzhou, China | Onsite video observation with 254 pedestrians | Illegal crossing behaviors at an unmarked, two-way, undivided, six-lane roadway | Safety margin | Multiple linear regression model | ◆ Pedestrians preferred crossing tentatively rather than waiting.<br>◆ Pedestrians preferred safe to short paths and crossed second half of the road faster.<br>◆ Pedestrians who were middle-aged, involved in a bigger group, looked at vehicles more often before crossing, and interacted with buses rather than private cars had a larger safety margin. |
| Cherry et al. (2012) | Kunming, China | Onsite video observation with 330 pedestrians | A marked, unprotected, two-way, eight-lane crosswalk | Time gap accepted, evasive actions taken by drivers | Probit regression model | ◆ The average accepted gap is 8.8 s and the average rejected gap is 5.3 s.<br>◆ Probability of accepting a gap was positively correlated with gap size and waiting time, but negatively correlated with vehicle-approaching speed. |
| Lobjois et al. (2013) | Caen, France | Virtual reality experiment with 67 participants | A single-lane urban street | Time gap accepted, crossing initiation time, and safety margin | Logistic regression model | ◆ Crossing behavior of the elderly was similar to that of younger counterparts on a single-lane, one-way street.<br>◆ Accepted time gap decreased when waiting time and number of passing vehicles increased. |
| Dommes et al. (2014) | Versailles, France | Virtual reality experiment with 84 participants | A marked, unprotected, two-way, two-lane midblock crosswalk | Time gap accepted, waiting time, crossing time, and safety margin | Logistic regression model | ◆ Old participants crossed more slowly and had smaller safety margins.<br>◆ Old participants made decisions mainly based on the gap available in the near lane while neglecting the far lane. |
| Koh and Wong (2014) | Singapore | Onsite video observation with 188 rejected and 104 accepted gaps, respectively | Illegal crossing behaviors at the stretch of seven signalized intersections | Time gap rejected and accepted | Logistic regression model | ◆ Averagely, a larger time gap accepted by violators was observed at the near-end (6.3 s) than the far-end crossings (5.2 s).<br>◆ For one-second increase in the time gap, the odds of illegal crossing increased 1.4 times for the near-end and 1.7 times for the far-end crossings, respectively. |
| Liu and Tung (2014) | Taiwan | Pre-recorded real-life video-based survey with 16 young and 16 elderly participants | An unprotected, two-lane crosswalk with one human-driven vehicle approached during midday and dusk | Walking time, time gap accepted, safety margin, confidence level, and walking strategy | Logistic regression model | ◆ Pedestrians made crossing decisions based mainly on the distance gap.<br>◆ The remining time to cross the road was longer and pedestrians adopted a walk-quickly strategy at dusk.<br>◆ The elderly didn't realize the decline of walking ability and made the same crossing decision with the younger, resulting in a smaller safety margin. |
| Petzoldt (2014) | Chemnitz, Germany | Video-based survey with 53 undergraduate students in Experiment I and 44 participants in Experiment II | Experiment I: a virtual video of an unprotected midblock with a single lane. Experiment II: a real-life video of a taxiway of an aviation airport. | Time gap accepted and time-to-arrival estimated by participants | Logistic regression model | ◆ Participants tended to accept a smaller time gap when the vehicle was approaching at a higher speed.<br>◆ The younger accepted a smaller time gap.<br>◆ A higher approaching speed resulted in a higher time-to-arrival estimate.<br>◆ The effect of vehicle-approaching speed decreased when using time-to-arrival estimate to predict gap acceptance. |
| Demiroz et al. (2015) | Izmir, Turkey | Onsite video observation with 454 jaywalkers | Illegal crossing behaviors at four overpass locations | Crossing speed, time gap, and safety margin | Descriptive statistics | ◆ About 46% of the pedestrians did not use the overpass and made illegal crossing.<br>◆ Pedestrians accepted larger time gaps on roads with a higher speed limit.<br>◆ Safety margin increased as the increase of pedestrian age and speed limit.<br>◆ Non-compliant pedestrians were aware of the danger of crossing illegally instead of using overpass. |
| Kadali et al. (2015) | Mumbai, India | Onsite video observation with 384 pedestrians | An unprotected midblock crosswalk at a divided six-lane urban street | Size of accepted time gap | Multiple linear regression and ANN models | ◆ ANN had better prediction performance while linear regression model produced more interpretable results.<br>◆ Elderly pedestrians were more likely to select larger time gaps.<br>◆ The accepted gap size decreased with the increase in waiting time, pedestrian group size, attempts made by the pedestrian to cross, and rolling behavior. |
| Onelcin and Alver (2015) | Izmir, Turkey | Onsite video observation with 444 jaywalkers | Illegal crossing behaviors at six two-lane signalized intersections | Distance gap, crossing time, and safety margin | Descriptive statistics | ◆ Vehicle-approaching speed had the most significant effect on decision to cross and crossing time.<br>◆ Pedestrian gender and age did not reveal a significant association with crossing time and distance-gap accepted. |
| Pawar and Patil (2015) | Kolhapur and Mumbai, India | Onsite video observation with 1107 gaps for 140 pedestrians | Two uncontrolled, marked midblock crosswalks on four-lane divided arterials | Time/distance gap accepted and rejected | Logistic regression model | ◆ The probability of accepted a spatial gap decreased with the increase in vehicle-approaching speeds and vehicle sizes.<br>◆ The $50^{th}$ percentile gap accepted varied from 4.1 to 4.8 s and from 67 to 79 m.<br>◆ The $85^{th}$ percentile gap accepted varied from 5.0 s to 5.8 s and from 82 to 95 m. |
| Pawar and Patil (2016) | Kolhapur and Mumbai, India | Onsite video observation with 1107 gaps for 140 pedestrians | Two uncontrolled, marked midblock crosswalks on four-lane divided arterials | Critical gap | Deterministic and probabilistic methods | ◆ Time and distance gaps followed the lognormal distribution.<br>◆ Critical gaps estimated by the deterministic methods were smaller than those of probabilistic methods.<br>◆ Critical gaps varied between 3.6–4.3 s and 60–74 m. |
| Naser et al. (2017) | Kuala Lumpur, Malaysia | Onsite video observation with unknown sample sizes | An uncontrolled midblock crosswalk | Time gap accepted | Multiple linear regression and logistic regression models | ◆ The elderly accepted a larger time gap, while the rolling gap crossing was associated with a smaller gap accepted.<br>◆ The probability of gap-acceptance decreased with smaller gap sizes, higher vehicle speeds, and larger vehicles. |
| Alver and Onelcin (2018) | Izmir, Turkey | Onsite video observation with 377 jaywalkers | Illegal crossing behaviors at two overpass locations | Time gap accepted, critical gap, and safety margin | Logistic regression model | ◆ The average accepted gap was 7.22 s.<br>◆ The critical gap varied from 3 to 3.3 s in the near-end lane.<br>◆ The average safety margin was 7.08 s. The younger who crossed individually and without items had a lower safety margin. |
| Shaaban et al. (2018) | Doha, Qatar | Onsite video observation with 2766 pedestrians | Illegal crossing behaviors on a six-lane divided arterial road | Waiting time, crossing path, crossing time, and crossing distance | Multiple linear regression model | ◆ The male was likely to cross illegally.<br>◆ The waiting time increased when pedestrians crossed from the curb, another pedestrian crossed from the opposite side, and with the absence of vehicles. |
| Sobhani and Farooq (2018) | Montreal and Toronto, Canada | VR experiment with 42 participants | Distracted crossing behaviors on a single-lane crosswalk at a unsignalized intersection | Crossing duration, waiting time, time-to-collision, and post-encroachment-tim e | Multinomial logit model | ◆ Distracted participants waited longer and walked faster to cross the street.<br>◆ The female had more dangerous crossing behaviors in distracted conditions.<br>◆ The LED light improved the safety of distracted pedestrians and enhanced the successful crossing rate. |
| Tapiro et al. (2018) | Beer-Sheva, Israel | VR experiment with 52 participants | Distracted crossing behaviors on a two-way, two-lane street | Time gap accepted, safety margin, crossing initiation time, and response time | Linear mixed model | ◆ Distracted participants chose smaller crossing gaps, took more time to make crossing decisions, were slower to respond to the crossing opportunity, and allocated less visual attention to roads.<br>◆ Visual distractions affected pedestrian behaviors more than the auditory. |
| Zhang et al. (2018) | Wuhan, China | Onsite video observation with unknown samples | Five uncontrolled, marked, two-way, six-lane midblock crosswalks | Crossing strategy, time gap accepted, and post-encroachment time | Multinomial logit regression model | ◆ The share of pedestrian selecting the single-stage, two-stage, and rolling-crossing strategy was 10.7%, 23.2%, and 66.1%, respectively.<br>◆ Female pedestrians were more likely to select the rolling-crossing strategy. |
| Avinash et al. (2019) | India | Onsite video observation with 4885 pedestrians | Four unprotected, undivided, four-lane midblock crosswalks | Safety margin | Multiple linear regression model | ◆ Safety margin increased with increase in crossing speed, available vehicle gap, and pedestrian group size, whereas decreased with pedestrian rolling behavior.<br>◆ Elderly and child pedestrians had a lower safety margin.<br>◆ Pedestrians were more likely to maintain a marginal gap for light vehicles. |
| Morrongiello et al. (2019) | Guelph, Canada | VR experiment with 86 child participants | A two-way, two-lane urban street | Time-gap accepted, waiting time, and proportion of being hit | Descriptive statistics | ◆ Children held similar views about crossing streets as they believed their peers held, and their norms were reflected by their recent crossing behaviors.<br>◆ Children who reported more risky crossing behaviors in the past few weeks selected a smaller time gap, waited less time, and experienced more hits in VR. |
| Zhang et al. (2019) | Wuhan, China | Onsite video observation with 3427 pedestrians | Twelve uncontrolled, marked midblock crosswalks | Safety margin | Ordered probit model | ◆ As the number of traffic lanes increased, the proportion of pedestrians adopting rolling gap crossing mode, crossing the street with others, and changing the speed or path increased accordingly.<br>◆ The number of pedestrian–vehicle conflicts at two-way six-lane crosswalks is 5.96 times higher than that of two-lane crosswalks, and 2.04 times higher than that of four-lane crosswalks. |
| Zhao et al. (2019) | Shanghai, China | On-site video observation with 11,500 pedestrians | Thirteen unprotected midblock crosswalks | Time gap accepted | Logistic regression model | ◆ The probability of gap acceptance increased with larger gaps, shorter crossing distance, and longer waiting time.<br>◆ Pedestrians waiting at the median were more likely to accept smaller time gaps. |
| Kadali and Vedagiri (2020) | Mumbai, India | On-site video observation with 46,170 accepted/rejected gaps | Eight uncontrolled midblock crosswalks | Time gap accepted | Logistic regression model | ◆ Elderly pedestrians, larger pedestrian groups, approaching of larger vehicles, higher vehicle speeds, smaller gap sizes, and wider crosswalks were associated with a lower probability of gap acceptance. |
| Sheykhfard and Haghighi (2020) | Babol, Iran | On-site video observations and in-vehicle video observations with 27 drivers | Two roads in urban and two road outskirt areas | Time gap accepted | Multiple linear regression and logistic regression models | ◆ Female pedestrians, larger vehicles, larger pedestrian group sizes, and higher vehicle speeds were associated with a lager time gap accepted, while those crossing immediately without waiting likely accepted a smaller time gap.<br>◆ The probability of gap acceptance increased with lower vehicle speeds, larger distance gaps, larger pedestrian group sizes, and longer waiting time. |
| Tapiro et al. (2020) | Beer-Sheva, Israel | VR experiment with 83 participants | Distracted crossing behaviors on a two-way, two-lane street | Safety margin, missed opportunities, and reaction time | Linear mixed model | ◆ Participants missed more opportunities to cross the road when exposed to more cluttered road environments.<br>◆ Children had a wider spread of gazes across the scene when the environment was highly loaded. |
| Vasudevan et al. (2020) | Kanpur, India | On-site video observation with 2401 gaps | Six uncontrolled, unmarked intersections | Critical gap and time gap accepted | Logistic regression model | ◆ Critical gap increased with the elderly, distracted pedestrians, larger vehicles, vehicle platoon, and longer waiting time.<br>◆ The probability of gap acceptance increased with a larger gap size and in front of vehicle platoon, but decreased with wider crossings, longer waiting time, and the approaching of larger vehicles. |
| Zafri et al. (2020) | Dhaka, Bangladesh | On-site video observation with 546 pedestrians | Rolling gap crossing behaviors on crosswalks at six intersections | Decision to adopt rolling gap crossing strategy | Logistic regression model | ◆ Uncontrolled intersections, smaller available gaps, heavy traffic flows, younger pedestrians, larger pedestrian group sizes, and not using crosswalks were associated with a higher likelihood of rolling gap crossing behaviors. |
| Alver et al. (2021) | Izmir, Turkey | On-site video observation with 498 pedestrians | Two unprotected, unmarked midblock crosswalks | Time gap accepted, critical gap, and crossing speed | Multiple linear regression model | ◆ The critical gap was 4.1 s in Bornova, and 6.2 s and 5.7 s for the first land and second lane, respectively, in Sirinyer.<br>◆ Pedestrians accepted smaller gaps where a physical barrier was built and when the approaching vehicle was a private car. |
| Leung et al. (2021) | Hong Kong, China | Pre-recorded real-life video-based survey with 906 child participants | An unprotected, one-way, two-lane midblock crosswalk | Reaction time and safety margin | Logistic generalized estimating equations | ◆ Children had difficulties in making safe judgment for vehicles approaching at over 30 km/h from the offside lane.<br>◆ Visual distractions were associated with poorer road crossing decisions. |
| Soares et al. (2021) | Portugal | VR experiment with 30 participants | Two unprotected, marked, one-way, two-lane midblock crosswalks | Percentage of crossings, response time, and time-to-arrival | Descriptive statistics | ◆ Lower speeds and higher distances leaded to higher percentages of crossings and longer response time.<br>◆ Participants' crossing decision was based on their visual perception of the approaching vehicle, particularly its speed and distance. |
| Soathong et al. (2021) | Auckland, New Zealand | On-site questionnaire survey with 400 participants | Risky and illegal crossing behaviors at four unprotected mid-blocks without pedestrian facilities | Intention to cross | Theory of planned behavior and structural equation model | • Intention to cross was mainly driven by habit and attitude.<br>• Some pedestrians internalized the belief that risky crossing behavior was an acceptable act in society.<br>• The female was highly influenced by their attitude while the male was influenced by their friends' perceptions. |
| Pawar and Yadav (2022) | Mumbai and Kolhapur, India | On-site video observation with 1107 pedestrians | Two uncontrolled midblock crosswalks | Time gap and distance gap accepted | Logistic regression model | • The probability of gap acceptance decreased with higher vehicle speeds, larger vehicles, smaller distance gaps, longer waiting time, and when pedestrians walked alone. |
| Wang et al. (2022) | Nantong, China | VR experiment with 193 valid participants | A marked, two-lane, two-way crosswalk in an uncontrolled intersection in urban areas | Start delay, missed opportunities, and dangerous crossing | Hierarchical multiple linear regression model | • Pedestrians low in sensation seeking missed more opportunities to cross.<br>• Harder traffic conditions and missed opportunities to cross were associated with more children's dangerous crossings.<br>• Those higher in sensation seeking, facing harder traffic, and with longer start delays had more dangerous crossings. |
| Angulo et al. (2023) | Virginia, US | On-site video observation with 153 pedestrians and VR experiment with 49 participants | A marked, unprotected midblock crosswalk on a two-way, two-lane urban street | Time gap accepted and crossing speed | Descriptive statistics | • The critical gap for real-world and VR was 5.12 s and 7 s, respectively.<br>• Pedestrians in the VR were more conservative, waiting longer and selecting larger time gaps.<br>• There was no significant difference in crossing speed between the VR and real-world. |
| Figueroa-Medina et al. (2023) | Mayagüez, Puerto, Rico | VR experiment with 48 participants | A marked, one-lane and one marked, two-lane, unprotected midblock crosswalk on an urban street | Time gap accepted, walking speed, and crossing success rate | Multiple linear and logistic regression models | • Participants, on average, watched about five vehicle gaps before crossing the street and accepted a gap of 4.5 s.<br>• Elderly participants on two-lane streets tended to accept larger time gaps.<br>• The probability of being hit increased with the elderly and higher vehicle speeds, but decreased with larger time gaps and higher walking speeds. |
| Kalantari et al. (2023) | Leeds, UK | Distributed simulator experiment with 64 participants | Two marked and two unmarked midblock crosswalks on a two-way urban street | Whether the pedestrian crossed first than the driver | Mixed-effect binary logit regression model | • Pedestrians tended to cross first at larger time gaps and on marked crosswalks.<br>• Pedestrians who had waited for longer time were less likely to cross first.<br>• Pedestrian decisions were affected mostly by vehicle kinematics.<br>• Personality traits showed limited effects on interaction at unmarked crossings. |
| Shen et al. (2023) | Shenyang, China | Pre-recorded real-life video-based survey with 60 child participants | Several marked, unprotected, three-lane crosswalks in the suburb areas | Dangerous crossing, safe crossing rate, and time-to-contact | Descriptive statistics | ◆ Six-year-old pedestrians exhibited lower safety in both behavioral and eye-tracking dimensions.<br>◆ The Chinese version of the Pedestrian Behavior Scale was positively associated with time-to-contact. |
| Zafri (2023) | Dhaka, Bangladesh | On-site video observation with 546 pedestrians | Six busy intersections | Crossing the road by walking or running | Association rules mining | ◆ Running was strongly associated with uncontrolled intersections, wide roads, narrow medians, male and younger pedestrians, rolling gap crossing, crossing alone, and crossing in front of light and faster vehicles. |
| Zhang et al. (2024) | Nanjing, China | On-site video observation with 319 jaywalkers | Illegal crossing behaviors at three two-way, two-lane midblock without crossing facilities | Sequential crossing decisions | Grouped random parameters multinomial logit model | ◆ Jaywalkers who previously made rushing decisions tended to continue making rushing decisions.<br>◆ Conservative vehicle status, such as decelerating, stopping, or swerving, were associated with a higher probability of pedestrian rushing decisions. |
| Bennett et al. (2025) | Melbourne and Sydney, Australia | VR experiment with 32 children and 44 adults | 77 video clips pre-recorded at 24 locations in real-world environments | Hazard perception and gap acceptance | Descriptive statistics | ◆ Adults responded more often within the hazard and gap windows, had significantly faster response times, and accurately identified hazards more often than children. |

1

**Appendix B**

**Table A2.** A summary of representative studies on motivational factors affecting pedestrian crossing behaviors when interacting with AVs.

| Study | Region | Experiment | Target variable | Motivational variables | Analysis method | Major findings |
|---|---|---|---|---|---|---|
| Deb et al. (2017b) | US | Online questionnaire survey with 482 participants | Intention to cross the road in front of AVs | Behavioral tendency, personal innovativeness, and pedestrian receptivity toward AVs | Principal component analysis, confirmatory factor analysis, and ordinal logit regression model | ◆ Intention to cross the road in front of AVs was significantly predicted by safety and interaction scores, but not by the compatibility score.<br>◆ The male, younger, urban residents, and those with higher personal innovativeness were more receptive toward AVs.<br>◆ Pedestrians who showed positive behaviors believed that AVs would improve overall traffic safety, while those with higher scores in violations, lapses, and aggressive behaviors felt more confidence to cross in front of AVs. |
| Deb et al. (2018) | Starkville, US | VR experiment with 30 participants | Crossing behaviors in front of AVs with different external features at a four-way, two-lane, and unsignalized intersection | Behavioral tendency, personal innovativeness, and pedestrian receptivity toward AVs | Linear regression model | ◆ Pedestrians' receptivity toward AVs significantly increased with the addition of external features.<br>◆ Pedestrians who often committed errors or showed aggressive behaviors rated AVs poorly, took longer time to cross the road, and waited less before crossing.<br>◆ Pedestrians who often violated traffic rule were more cautious and appreciated the inclusion of external features. |
| Rodríguez Palmeiro et al. (2018) | Delft, Netherlands | A Wizard of Oz experiment with 24 participants | Intention to cross a two-way, six-lane, and unmarked road in front of AVs or human-driven vehicles | Sensation seeking and trust in AVs | Descriptive statistics | ◆ No significant differences in critical gap and self-reported stress were observed between AVs and human-driven vehicles.<br>◆ The male accepted shorter gaps and had higher sensation seeking scores.<br>◆ There were no significant differences between the size of critical gap and self-reported stress levels, trust in AVs, and sensation seeking. |
| Dey et al. (2019) | Delft, Netherlands | Pre-recorded videos of ghost drivers with 50 participants | Willingness to cross an unmarked road in a campus in front of AVs or human-driven vehicles | Overall assertiveness and knowledge of AVs | Descriptive statistics | ◆ Vehicle automation, assertiveness, and knowledge of AVs did not have a significant influence on pedestrians' willingness to cross.<br>◆ The yielding behavior of vehicles played a dominant role in pedestrians' decision to cross. |
| Nuñez Velasco et al. (2019) | Delft, Netherlands | VR experiment with 55 participants | Intention to cross the road in front of AVs with different physical appearance and external human-machine interfaces | Behavioral tendency, perceived behavioral control, and trust in automation | Mixed binomial logit regression model | ◆ The presence of zebra crossings and larger gaps increase crossing intentions.<br>◆ Pedestrians who recognized the vehicle as an AV had a lower intention to cross.<br>◆ Pedestrians who recognized the vehicle as automated showed a higher level of trust in AVs.<br>◆ A significantly positive relationship was found between crossing intention and perceived behavioral control. |
| Woodman et al. (2019) | Warwick, UK | VR experiment with 28 participants | Intention to cross in front of a platoon with four AVs on shared space or on a single-lane road | Perceived safety | Descriptive statistics | ◆ Pedestrians were more likely to accept a smaller time gap in the shared environments, although lower perceived safety was reported. |
| Deb et al. (2020) | Starkville, US | VR experiment with 24 participants | Crossing behaviors in front of AVs with different operator states and external features at a four-way, two-lane, and unsignalized intersection | Personal innovativeness and pedestrian receptivity toward AVs | Descriptive statistics | ◆ Pedestrians preferred both "walk" in text and verbal message saying "safe to cross" as clear and comfortable interfaces.<br>◆ Pedestrians trusted an AV without an operator over the AV with a distracted operator.<br>◆ The older found the external interfaces more helpful and those with higher innovativeness rated the feature ideas with higher scores. |
| Rad et al. (2020) | Delft, Netherlands | Simulation experiment with 60 participants | Intention to cross the one-way, one lane road in front of AVs or human-driven vehicles | Behavioral tendency, risk perception, and attitudes toward AVs | Principal component analysis and generalized linear mixed model | ◆ Distance to approaching vehicles and the presence of zebra crossings were two most important to predict pedestrians' intention to cross.<br>◆ Pedestrians had a significantly higher intention to cross the road in front of AVs than human-driven vehicles.<br>◆ Pedestrians who were familiar with AVs, had more trusts in AVs, and reported more violation behaviors were more likely to cross the road in front of AVs. |
| Bindschädel et al. (2021) | Weissach, Germany | VR experiment with 51 participants | Crossing behaviors in front of AVs with and without external human-machine interfaces on a one-way, one-lane, and unmarked midblock crosswalk | Perceived safety | Linear mixed model | ◆ Participants crossed the road earlier and felt safer when encountering AVs with external human-machine interfaces.<br>◆ In situations in which only some of the AVs were equipped with external interfaces, participants became more conservative for encounters without external human-machine interfaces.<br>◆ Subjective safety feeling was significantly predicted from actual crossing behaviors. |
| Dommes et al. (2021) | Versailles, France | VR experiment with 30 younger and 30 older participants | Crossing behaviors in front of mixed traffic with AVs and human-driven vehicles on a two-way, two-lane, and unmarked midblock crosswalk | Pedestrian receptivity toward AVs | Linear mixed model | ◆ Participants hesitated to cross in front of an AV, with later initiations and longer crossing time.<br>◆ Participants tended to cross the street at shorter gaps, when AVs gave way in the near lane while conventional vehicles were approaching in the far lane.<br>◆ No significant differences in receptivity toward AVs were found between the younger and older participants. |
| Nuñez Velasco et al. (2021) | Leeds, UK | VR experiment with 20 participants | Crossing behaviors in front of AVs with different driver conditions on a single-lane one-way road | Trust in AVs, perceived behavioral control, and perceived risk | Mixed effect linear regression model | ◆ Time gaps and the yielding behavior of AVs were the most important factors affecting pedestrians' crossing behaviors.<br>◆ Those who thought the vehicle were AVs were less likely to cross.<br>◆ Trust in AVs and perceived risk were not significantly associated with crossing decisions.<br>◆ Those with higher perceived behavioral controls were more likely to cross.<br>◆ Perceived behavioral control was higher and perceived risk was lower when the driver in AVs appeared attentive. |
| Kwon et al. (2022) | Ulsan and Seoul, South Korea | VR experiment with 200 young participants | Crossing behaviors at a four-way, unsignalized intersection in a residential block | Risk perception | Random-effect linear regression and hierarchical SEM | ◆ Risk perception was higher with barriers of visibility, without geometric patterns, without pavement markings, in nighttime, and in the shared street.<br>◆ Participants with a higher perceived risk took longer to start walking and tended to walk in haste while crossing the road.<br>◆ Risk perception mediates the relationship between built environment and crossing behaviors. |
| Zhao et al. (2022) | Australia | Online questionnaire survey with 493 participants | Intention to cross the road in risky situations in front of AVs or human-driven vehicles | Theory of planned behavior, perceived risk, and trust | Hierarchical multiple linear regression model | ◆ Pedestrians had a significantly higher intention to cross the road in front of AVs than human-driven vehicles.<br>◆ Pedestrians reported lower risks and greater trusts toward road-crossing in front of AVs.<br>◆ Attitude, subjective norm, and perceived behavioral control were significant predictors of intentions to engage in risky road-crossing behaviors. |
| Feng et al. (2024) | China | Virtual video-based online questionnaire survey with 589 participants | Intention to cross a two-way, two-lane, unsignalized, and marked midblock crosswalk in front of AVs or human-driven vehicles | Risk perception and trust | Integrated choice and latent variable model | ◆ An increase in vehicle approaching-speed and a decrease in distance gap increased pedestrians' tendency to wait.<br>◆ Middle-aged pedestrians and those with higher perceived risks were more conservative regarding road-crossing decisions.<br>◆ Higher levels of trust in AV improved pedestrians' willingness to cross. |

## Appendix C

**Table A3.** Items and variables included in the questionnaire survey.

| Variables | Items |
|---|---|
| Violations | I cross the street even though the pedestrian light is red.<br>I cross diagonally to save time.<br>I cross outside the pedestrian crossing even if there is one (crosswalk) less than 50 meters away.<br>I take passageways forbidden to pedestrians to save time. |
| Errors | I cross between vehicles stopped on the roadway in traffic jams.<br>I cross even if vehicles are coming because I think they will stop for me.<br>I walk on cycling paths when I could walk on the sidewalk.<br>I run across the street without looking because I am in a hurry. |
| Lapses | I realize that I have crossed several streets and intersections without paying attention to traffic.<br>I forget to look before crossing because I am thinking about something else.<br>I cross without looking because I am talking with someone.<br>I forget to look before crossing because I want to join someone on the sidewalk on the other side. |
| Aggressive behaviors | I get angry with another road user (pedestrian, driver, cyclist, etc.), and I yell at him.<br>I cross very slowly to annoy a driver.<br>I get angry with another road user (pedestrian, driver, cyclist, etc.), and I make a hand gesture.<br>I got angry with a driver and hit his vehicle. |
| Positive behaviors | I thank the driver who stopped to let me cross.<br>When I am accompanied by other pedestrians, I walk in a single file on narrow sidewalks so as not to bother the pedestrians I meet.<br>I walk on the right-hand side of the sidewalk so as not to bother the pedestrians I meet.<br>I let a car go by, even if I have the right-of-way, if there is no other vehicle behind it. |
| Understanding of trucks | I understand the main uses and functions of trucks.<br>I am aware of the advantages and disadvantages of truck transportation.<br>I am familiar with the driving and operating skills of trucks. |
| | I am knowledgeable about the relevant traffic laws and regulations for truck transportation. |
| Trust in safety of trucks | I believe that trucks pose a significant safety risk to pedestrians when driving on urban roads.<br>I am concerned that the size of trucks may limit the drivers' visibility, increasing the likelihood of accidents.<br>Compared to regular cars, I think trucks are more difficult for people to avoid on the road |
| Understanding of autonomous driving technology | I was aware of the uses and functions of autonomous vehicles.<br>I understood how autonomous vehicles operate and their working principles.<br>I know about the history and prospects of the autonomous vehicle industry. |
| Trust in safety about autonomous driving technology | I believe that autonomous vehicles can detect pedestrians well and avoid collisions.<br>I think the risk of accidents when autonomous vehicles interact with pedestrians is low.<br>I believe that autonomous driving technology can effectively protect the safety of pedestrians in emergency situations. |
| Trust in safety of autonomous trucks | I would feel safer crossing the road in front of an autonomous truck.<br>The presence of autonomous trucks will make the entire road traffic safer.<br>I would feel more at ease if my family crosses the road on streets where autonomous trucks are present.<br>Crossing the road in front of an autonomous truck, I would feel more comfortable doing other things (such as using my phone, talking with companions).<br>When encountering an autonomous truck, I would spend less effort observing the surroundings and crossing the road.<br>Autonomous trucks will be able to interact effectively with other vehicles and pedestrians. |
| Risk perception | I think the danger of crossing the road in the situation shown in the video is very high.<br>If I were to cross the road in the situation depicted in the video, the consequences of an accident would be very serious.<br>My vigilance increases when an autonomous truck is approaching.<br>I feel that the autonomous truck in the video might lose control and pose a danger. |
| Gap acceptance | Please press the button of handheld remote at the last moment that you feel safe to cross in the virtual environment |

## Appendix D

**Table A4.** Descriptive statistics of accepted gap size (unit: seconds).

| | **2 s** | **3 s** | **4 s** | **5 s** | **6 s** | **7 s** | **8 s** | **9 s** |
|---|---|---|---|---|---|---|---|---|
| Frequency | 5 | 56 | 86 | 211 | 140 | 44 | 50 | 11 |
| Proportion | 0.82% | 9.29% | 14.26% | 34.99% | 23.22% | 7.30% | 8.29% | 1.82% |
| Mean | | | | 5.35 s | | | | |
| Standard deviation | | | | 1.43 s | | | | |

## Appendix E

**Fig. A1.** Estimation results of measurement model.

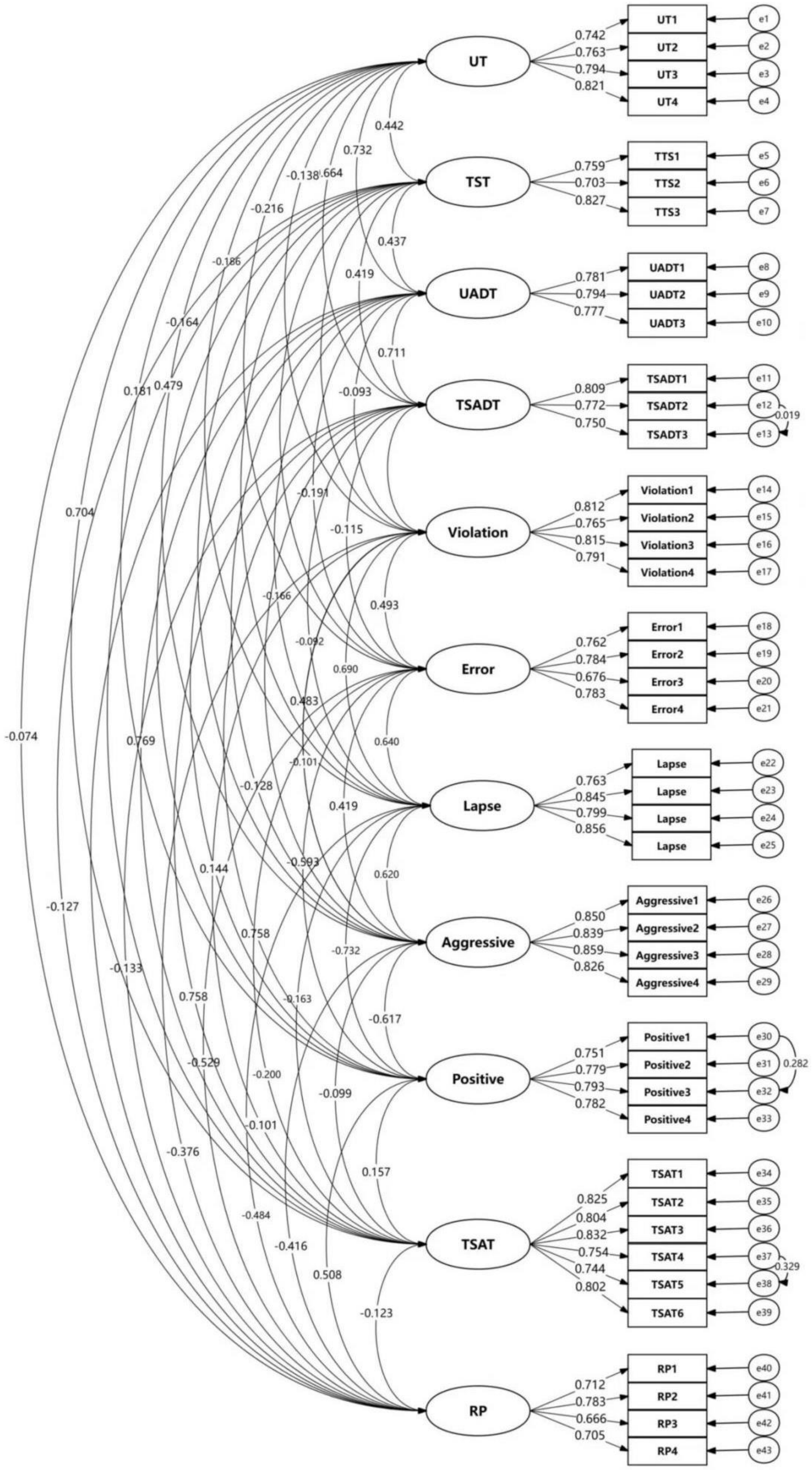

**Appendix F**

**Table A5.** Results of discriminant validity tests (UT: understanding of trucks; TST: trust in the safety of trucks; UADT: understanding of automated driving technology; TSADT: trust in the safety of automated driving technology; TSAT: trust in the safety of automated trucks).

| | UT | TST | UADT | TSADT | Violations | Errors | Lapses | Aggressive | Positive | TSAT | Risk perception |
|---|---|---|---|---|---|---|---|---|---|---|---|
| UT | 0.78 | | | | | | | | | | |
| TST | 0.44 | 0.77 | | | | | | | | | |
| UADT | 0.73 | 0.44 | 0.78 | | | | | | | | |
| TSADT | 0.66 | 0.42 | 0.71 | 0.78 | | | | | | | |
| Violations | –0.14 | 0.03 | –0.09 | –0.02 | 0.80 | | | | | | |
| Errors | –0.22 | –0.08 | –0.19 | –0.12 | 0.49 | 0.75 | | | | | |
| Lapses | –0.19 | 0.03 | –0.17 | –0.09 | 0.69 | 0.64 | 0.82 | | | | |
| Aggressive | –0.16 | –0.03 | –0.13 | –0.05 | 0.48 | 0.42 | 0.62 | 0.84 | | | |
| Positive | 0.18 | 0.04 | 0.14 | 0.12 | –0.64 | –0.59 | –0.73 | –0.62 | 0.78 | | |
| TSAT | 0.70 | 0.48 | 0.77 | 0.76 | –0.10 | –0.20 | –0.16 | –0.10 | 0.16 | 0.79 | |
| Risk perception | –0.07 | –0.13 | –0.07 | –0.13 | –0.38 | –0.53 | –0.48 | –0.42 | 0.51 | –0.12 | 0.72 |